\documentclass[iop]{emulateapj}

\begin{document}

\title{Hot-Dust (690\,K) Luminosity Density and its Evolution in the last 7.5 Gyr}

\author{H. Messias\altaffilmark{1,2}, B. Mobasher\altaffilmark{3}, J. Afonso\altaffilmark{1,4}}
\email{hmessias@oal.ul.pt}

\altaffiltext{1}{Centro de Astronomia e Astrof\'{\i}sica da Universidade de Lisboa,  Observat\'{o}rio Astron\'{o}mico de Lisboa, Tapada da Ajuda, 1349-018 Lisbon, Portugal}
\altaffiltext{2}{Departamento de Astronom\'ia, Av. Esteban Iturra 6to piso, Facultad de Ciencias F\'isicas y Matem\'aticas, Universidad de Concepci\'on, Chile.}
\altaffiltext{3}{Department of Physics and Astronomy, University of California, 900 University Ave., Riverside, CA 92521, USA.}
\altaffiltext{4}{Department of Physics, Faculty of Sciences, University of Lisbon, Campo Grande, 1749-016 Lisbon, Portugal}

\begin{abstract}
\noindent We study the contribution of hot-dust to the luminosity density of galaxies and its evolution with cosmic time. Using the \emph{Spitzer}-IRAC data over an area of 1.8\,deg$^2$ covered by the COSMOS field, we estimate the contribution from hot-dust at rest-frame 4.2\,$\mu$m (from $0 \lesssim z \lesssim 0.2$ up to $0.5 \lesssim z \lesssim 0.9$). This wavelength corresponds to black-body temperature of $\sim690\,$K. The contribution due to stellar emission is estimated from the rest-frame 1.6\,$\mu$m luminosity (assumed to result from stellar emission alone) and subtracted from the mid-infrared luminosity of galaxies to measure hot-dust emission. In order to attempt the study of the 3.3\,$\mu$m-PAH feature, we use the rest-frame 4.2\,$\mu$m to infer the hot-dust flux at 3.3\,$\mu$m. This study is performed for different spectral types of galaxies: early-type, late-type, starburst, and IR-selected active galactic nuclei (AGN). We find that: (a) the decrease of the hot-dust luminosity density since $0.5 \lesssim z \lesssim 1$ is steeper (by at least 0.5\,dex) compared to that of the cold-dust, giving support to the scenario where galaxy obscuration increases with redshift, as already proposed in the literature; (b) hot-dust and PAH emission evolution seems to be correlated with stellar mass, where rest-frame 1.6\,$\mu$m luminous non-AGN galaxies (i.e., massive systems) show a stronger decrement (with decreasing redshift) in hot-dust and PAH emission than the less luminous (less massive) non-AGN galaxies; (c) despite comprising $\lesssim3\%$ of the total sample, AGN contribute as much as a third to the hot-dust luminosity density at $z<1$ and clearly dominate the bright-end of the total hot-dust Luminosity Density Function at $0.5\lesssim z\lesssim0.9$; (d) the average dust-to-total luminosity ratio increases with redshift, while PAH-to-total luminosity ratio remains fairly constant; (e) at $\rm{M}_{1.6}>-25$, the dust-to-total and PAH-to-total luminosity ratios increase with decreasing luminosity, but deeper data is required to confirm this result. Future study is necessary to enlighten further the characterization of the different spectral components at play in 2--5\,$\mu$m spectral regime.
\end{abstract}

\keywords{galaxies: high-redshift; galaxies: active; galaxies: starburst}

\section{Introduction}

Detailed knowledge of dust and its effect on the total bolometric luminosity of galaxies is essential for an unbiased study of the evolution of galaxies. For example, dust is often associated with star formation activity and, hence, any measurement of this requires correction for dust extinction \citep{Silva98,Buat05,Bouwens09}. Furthermore, dust is directly correlated with the metallicity of galaxies and their chemical evolution \citep[e.g.,][]{Calzetti07}. As dust is mainly produced by supernov\ae\ \citep{Rho08,Barlow10} or low/intermediate mass asymptotic giant branch (AGB) stars \citep{Gehrz89,FerrarottiGail06,Sargent10}, it also provides a clue towards studying stellar evolution in galaxies. Finally, the knowledge of the effect of dust at high redshifts is important in studies of early formation of galaxies, their star formation rates and mass assembly \citep[e.g.,][and references therein]{Hainline11}. 

Dust affects the output energy of galaxies by absorbing and re-processing their UV-to-optical light (produced by star forming regions or black-hole accretion), resulting in its re-emission mostly at far-infrared (FIR, $\sim70-500\,\mu$m throughout) wavelengths. Hence, the IR spectral regime (and the millimeter spectral regime at higher redshifts) has been most useful in unveiling the properties of dust in galaxies \citep[for a review, see][]{Hunt10}. However, much of this work \citep[e.g.,][]{Saunders90a,Saunders90b,Blain99,Scott02,Greve04,Greve08,Ivison05,Mortier05,Magnelli09,Clements10,Jacobs11} has relied on shallow data or small number statistics when compared to optical/near-infrared-based studies, even though stacking analysis have been used to lessen this problem \citep[e.g.,][]{CharyPope10}. While the FIR/mm spectral range is sensitive to the cold dust component (T$\lesssim$100\,K) mostly found in the inter-stellar medium (ISM), the hot dust (T$\sim$500--1500\,K) component, mostly emitting at $\sim$1--8\,$\mu$m, is nearer to the heating source, thus being the front-line medium to absorb direct UV/optical light and trace the most active regions in galaxies.

Here, we perform a statistical study of the contribution of hot-dust and 3.3\,$\mu$m Polycyclic Aromatic Hydrocarbon (PAH) feature to the observed spectral energy distributions (SEDs) of different populations of galaxies and their evolution with redshift. The results here can be used to quantify the amount of the UV/optical light obscuration in galaxies undergoing different degrees of star-formation or nuclear activity and its evolution with look-back time. this is done by isolating the hot-dust component in galaxies and performing a statistical measurement of its luminosity density. Such a study requires wide-area, multi-waveband data. Particularly important, is the availability of deep or medium-deep near- to mid-infrared (NIR, MIR) data (1--8\,$\mu$m) tracing the hot-dust regime. These requirements are achieved by the observations from the IR array camera \citep[IRAC,][]{Fazio04} on board the \textit{Spitzer Space Telescope} \citep[\textit{Spitzer},][]{Werner04} covering the Cosmic Evolution Survey \citep[COSMOS,][]{Scoville07}.

In Section~\ref{sec:samp}, the sample used in this study is described, and the redshift estimates are presented in Section~\ref{sec:reds}. In Section~\ref{sec:apcorr} we correct the aperture-to-total flux estimates as a function of redshift. Section~\ref{sec:estimdust} describes the method used to estimate dust and PAH emission, using SEDs for individual galaxies. The evolution of dust-to-total and PAH-to-total luminosity ratios are studied in Section~\ref{sec:d2t}. We present the dust and PAH luminosity density functions (LDFs), their dependence on galaxy types and their evolution with redshift in Section~\ref{sec:dlfs}. The dependency on galaxy type and redshift is also shown for the overall dust and PAH luminosity densities in Section~\ref{sec:ldz}. Finally, Section~\ref{sec:conc} summarizes the conclusions.

Throughout this paper we use the AB magnitude system\footnote{When necessary the following relations are used: ($K_s$, [3.6], [4.5], [5.8], [8.0])$_{AB}$ = ($K_s$, [3.6], [4.5], [5.8], [8.0])$_{Vega}$ + (1.841, 2.79, 3.26, 3.73, 4.40) from \citet{Roche03} and \emph{http://spider.ipac.caltech.edu/staff/gillian/cal.html}.}. A $\Lambda$CDM cosmology is assumed with H$_{0} = 70$\,km\,s$^{-1}$\,Mpc$^{-1}$, $\Omega_{M} = 0.3$, $\Omega_{\Lambda} = 0.7$.

\section{The sample} \label{sec:samp}

We use observations from the COSMOS field, covering an area of 1.8\,deg$^2$, with available multi-waveband data. The ultraviolet-to-IR coverage (0.15--8\,$\mu$m) is unique and there are, in total, 30 broad, intermediate, and narrow-band filters available \citep[Table~1 in][I09 henceforth]{Ilbert09}.

For the present study, we use both the $i^+$-band (\textit{Subaru} Telescope) and 3.6\,$\mu$m-selected catalogs described in I09. Our sample selection is based on flux cuts applied to the appropriate IRAC band tracing the rest-frame 3.3\,$\mu$m wavelength at a given redshift interval.

\subsection{Galaxy populations} \label{sec:gal}

We divide the galaxies in this study into early-type, late-type, starburst and active galactic nuclei (AGN), based on their SEDs. The non-AGN classification results from fitting galaxy SED templates to the observed galaxy SEDs as briefly described in Section~\ref{sec:reds} (see the detailed description in I09). For each galaxy, the best fit template reveals its spectral type.

The AGN in our sample are color-color selected with the following criteria: $K_s-[4.5]>0$ and $[4.5]-[8.0]>0$ \citep[KI,][]{Messias12}. Whenever the $K_s$-band is not available, sources are considered to be AGN when satisfying the \citet{Donley12} criterion or when brighter than a rest-frame $K_s$ absolute magnitude of M$_{K_s}=-26$ (for the 211 sources in our sample with unavailable $K_s$-band photometry, 55 are IR AGN). No other spectral regime (X-rays, optical, or radio) is adopted to identify AGN hosts. The goal of this study is to independently trace IR emission due to hot-dust, induced by different radiation mechanisms. However, X-rays/optical/radio-selected AGN may reveal little of the hot-dust emission induced by the AGN itself, with the IR emission being potentially dominated by stellar light or dust emission from the host galaxy. The IR-selection ensures that the hot-dust emission observed in AGN is mostly induced by nuclear activity.

\subsection{Tracing hot-dust emission} \label{sec:rfwl}

This study focuses on two adjacent spectral regimes at rest-frame wavelengths 3.3\,$\mu$m and 4.2\,$\mu$m. These are chosen so that hot-dust emission continuum is traced in both regimes (corresponding to black-body temperatures of $\sim880\,$K and $\sim690\,$K, respectively), as well as the known PAH feature at 3.3\,$\mu$m. By comparing the two adjacent spectral regimes, we aim to infer differences between hot-dust and PAH emissions and assess their evolution with cosmic-time.

Although PAHs are not actual dust particles, they are large molecules of Carbon rings (composed of $\sim$50 Carbon atoms) and Hydrogen, which act as small dust grains significantly blocking UV radiation, producing broad emission features in the IR SEDs of galaxies \citep[for a review, see ][]{Tielens11}. Also, PAHs comprise a non-negligible fraction of the Carbon existing in the Universe (5--10\,\%) and are believed to be closely related to star-formation activity \citep[][and references therein]{Tielens11}.

The 4.2\,$\mu$m spectral regime is free from PAH emission and traces only IR continuum due to dust heated by energetic radiation fields. Obscured star-formation as well as AGN activity can account for such emission \citep[][and references therein]{daCunha08,Nenkova08,HonigKishimoto10,Popescu11}.

\subsection{Sample selection} \label{sec:sampsel}

The final samples are first selected based on redshift criteria (Table~\ref{tab:zrange}) allowing the same rest-frame wavelength to be traced by each of the IRAC filters. The redshift intervals considered in this study are set by the target 3.3\,$\mu$m rest-frame wavelength and the widths of the IRAC filters. Table~\ref{tab:zrange} shows the adopted redshift intervals for this study, resulting from the specific redshifts where the 3.3\,$\mu$m wavelength enters or leaves the 50\% throughput limits of an IRAC filter. The rest-frame 4.2\,$\mu$m wavelength is then traced by the contiguous IRAC filter at longer-wavelengths relative to the band tracing rest-frame 3.3\,$\mu$m. As a result, the work is limited to the $z\lesssim1$ range, where rest-frame 4.2\,$\mu$m is still possible to observe with the IRAC filter-set. The redshift estimates for individual galaxies are described in Section~\ref{sec:reds}.

\begin{deluxetable}{crrrrr}
\tabletypesize{\normalsize}
\tablecaption{The adopted redshift ranges and equivalent observing bands for the sampling of the rest-frame 3.3\,$\mu$m.\label{tab:zrange}}
\tablewidth{0pt}
\tablehead{
\colhead{IRAC$_{\lambda}$} & \colhead{$z_{\rm LOW}$} & \colhead{$z_{\rm HIGH}$} & \colhead{Volume} & \multicolumn{2}{c}{Scale\tablenotemark{a}} \\
\colhead{} & \colhead{} & \colhead{} & \colhead{} & \colhead{1$.\!\!^{\prime\prime}$9} & \colhead{4$.\!\!^{\prime\prime}$1} \\
\colhead{[$\mu$m]} & \colhead{} & \colhead{} & \colhead{[$10^{6}$\,Mpc$^3$]} & \multicolumn{2}{c}{[kpc]}
}
\startdata
3.6 & 0.05 & 0.19 & 0.84 & 1.9--6.0 & 4.0--13.0 \\
4.5 & 0.21 & 0.52 & 1.25 & 6.5--11.8 & 14.1--25.5 \\
5.8 & 0.52 & 0.94 & 4.38 & 11.8--15.0 & 25.5--32.3 
\enddata
\tablenotetext{a}{The physical scale corresponding to the 1$.\!\!^{\prime\prime}$9 and 4$.\!\!^{\prime\prime}$1 apertures considered for IRAC photometry. These values come from Ned Wright's Cosmology calculator (http://www.astro.ucla.edu/$\sim$wright/CosmoCalc.html) by giving as input the adopted cosmology in this study. The study is limited to the $z<0.94$ range where rest-frame 4.2\,$\mu$m is possible to trace by the IRAC filter-set.\\}
\end{deluxetable}

\begin{figure}
\includegraphics[scale=0.33,angle=-90]{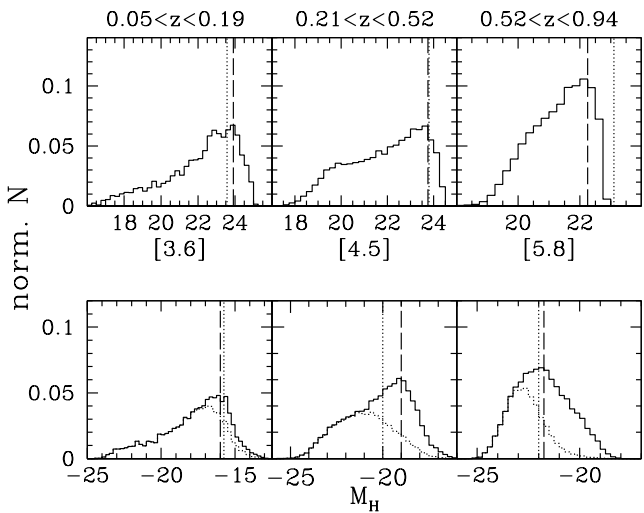}
\caption{Magnitude cuts to determine sample completeness. Different columns refer to different redshift intervals: $0.05<z<0.19$ (left); $0.21<z<0.52$ (middle); $0.52<z<0.94$ (right). The top panels refer to the apparent magnitude cut in the band tracing rest-frame 3.3\,$\mu$m: 3.6\,$\mu$m (left), 4.5\,$\mu$m (middle), 5.8\,$\mu$m (right). The dashed-line indicates the magnitude beyond which the magnitude distribution starts to drop, while the dotted-line indicates the magnitude beyond which the redshift completeness is smaller than 0.7 (Section~\ref{sec:reds}). The histograms are normalized to the sample size ($\rm{N_{BIN}/N_{TOT}}$): 4560 (left); 23915 (middle); 32265 (right). The bottom panels refer to the rest-frame 1.6\,$\mu$m absolute magnitude cut. The dashed-line indicates the magnitude beyond which the magnitude distribution starts to drop, while the dotted-line indicates the magnitude beyond which the hot-dust and PAH estimate completeness drops below 0.7. The histograms are normalized to the sample size (galaxies for which M$_H$ was possible to estimate): 4560 (left); 23759 (middle); 32265 (right).\label{fig:magcut}}
\end{figure}

We apply two completeness magnitude cuts to the resulting redshift selected samples. The first (upper panels in Figure~\ref{fig:magcut}) considers apparent magnitudes and is set as the brightest between the magnitude value at which the magnitude distribution starts to drop and the magnitude beyond which the fraction of sources with a redshift estimate (redshift completeness) is lower than 0.7 (Section~\ref{sec:reds}).

The second completeness magnitude cut (lower panels in Figure~\ref{fig:magcut}) considers rest-frame 1.6\,$\mu$m absolute magnitudes (assumed to be dominated by stellar light alone) and takes into account the fact that it is not always possible to simultaneously trace the rest-frame 3.3 and 4.2\,$\mu$m for one galaxy. This is driven by the fact that rest-frame 4.2\,$\mu$m is always traced by shallower data. This magnitude cut is set as the brightest between $\rm{M_H}$ value beyond which the fraction of sources with a hot-dust and PAH estimate drops below 0.7 and the value at which the magnitude distribution starts to drop.

Hence, we consider all $0.05<z<0.19$ sources with $[3.6]<23.56$ and $\rm{M_H<-16}$, all $0.21<z<0.52$ sources with $[4.5]<23.75$ and $\rm{M_H<-20}$, and all $0.52<z<0.94$ sources with $[5.8]<22.25$ and $\rm{M_H<-22}$. No source flagged photometrically (e.g., due to blending or bad pixels) was considered in this study, and no incompleteness correction regarding this rejection is attempted. The final number counts for different populations are listed in Table~\ref{tab:numb}.

\begin{deluxetable*}{crrrrr}
\tabletypesize{\normalsize}
\tablecaption{The numbers of each population with redshift.\label{tab:numb}}
\tablewidth{0pt}
\tablehead{
\colhead{$z_{\rm BIN}$} & \colhead{TOTAL} & \colhead{EARLY} & \colhead{LATE} & \colhead{STARB.} & \colhead{AGN}
}
\startdata
$0.05<z<0.19$ &  2954 &  514\,(17) &  341\,(12) & 2081\,(70) &  18\,(1) \\
$0.21<z<0.52$ & 11439 & 2383\,(21) & 3418\,(30) & 5545\,(48) &  93\,(1) \\
$0.52<z<0.94$ & 14435 & 3198\,(22) & 4470\,(31) & 6340\,(44) & 427\,(3)
\enddata
\tablecomments{Numbers in parenthesis give the fraction (in \%, approximated to unit) of the total population each population represents at each redshift interval.}
\end{deluxetable*}

\section{Redshift estimates} \label{sec:reds}

Whenever available, a spectroscopic redshift is assigned to galaxies. However, the bulk of the redshift estimates are photometric, based on the broad-to-narrow-band photometry coverage available.

Briefly, the photometric redshifts are measured using a SED fitting procedure using the \textit{Le Phare} code\footnote{http://www.cfht.hawaii.edu/$\sim$arnouts/LEPHARE/\\cfht\_lephare/lephare.html} (S. Arnouts \& O. Ilbert). The procedure applied on the COSMOS photometry data-set is described in I09. In the worst case scenario for our sample (at $z\sim1$ and $i^+\sim25$), the photometric redshift accuracy is expected to be $\sigma_{\Delta{z}/(1+z_{\rm spec})}\sim0.05$ (where $\Delta{z}=z_{\rm spec}-z_{\rm phot}$) and the percentage of outliers (sources with $|z_{\rm spec}-z_{\rm phot}|/(1+z_{\rm spec})>0.15$, denoted by $\eta$) is $\eta\sim20\%$ (I09).

\citet[][S11 henceforth]{Salvato09,Salvato11} computed photometric redshifts for XMM-\emph{Newton} and \emph{Chandra}-detected sources. Variability effects in X-ray AGN hosts are addressed before computing photometric redshifts, and SED templates accounting for AGN emission contribution are also adopted. This method provides photometric redshifts for the X-ray AGN sample with an excellent quality, reaching an accuracy of $\sigma_{\Delta z/(1+z_{\rm spec})}\sim0.015$ with an outlier fraction of $\eta\sim6\%$. One conclusion from this study was that below a given soft X-ray flux ($F_{(0.5-2\,{\rm keV})}=8\times10^{-15}\,{\rm erg\,s^{-1}}$), non-variable and/or morphologically extended X-ray sources do not require SED templates taking into account AGN emission. However, the recipe adopted in S11 should not be applied to X-ray-undetected sources, as intrinsically X-ray-luminous AGN may remain undetected due to high obscuration. Appendix~\ref{sec:zphagn} presents a detailed discussion of the photometric redshift quality for the IR-selected AGN sample (described in Section~\ref{sec:gal}).

Some sources in the sample will have more than one redshift estimate resulting from different methods. The final redshift priority sequence (from highest to lowest) is as follows:
\begin{itemize}
\item[-] spectroscopic redshifts with a probability larger than $90\%$ of being correct from the $z$COSMOS catalog, which considers only $i^+<22$ sources \citep{Lilly09};
\item[-] spectroscopic redshifts from \citet{Trump09}, \citet{Brusa10}, \citet{Fu10}, \citet{Kartaltepe10}, and \citet{Knobel12}, which provide additional spectroscopy for fainter sources;
\item[-] photometric redshifts from S11 for XMM-\emph{Newton} and \emph{Chandra} detected sources;
\item[-] photometric redshifts from I09 for non-IR-AGN sources;
\item[-] combination of photometric redshift solutions using non-AGN or AGN templates for IR-AGN sources as described in the Appendix~\ref{sec:zphagn}.
\end{itemize}

\begin{figure}
\plotone{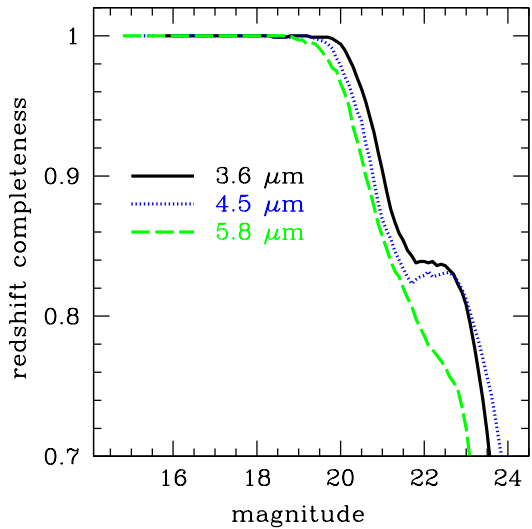}
\caption{Completeness of reliable redshift estimates (i.e., fraction of sources with $(z_{68\%}^{\rm up}-z_{68\%}^{\rm low})/(1+z_{\rm phot})<0.4$ and $i^+<25$) depending on source magnitude in each of the IRAC channels: 3.6\,$\mu$m (solid black line), 4.5\,$\mu$m (dotted blue line), and 5.8\,$\mu$m (dashed green line).\label{fig:zincomp}}
\end{figure}

\begin{figure}
\plotone{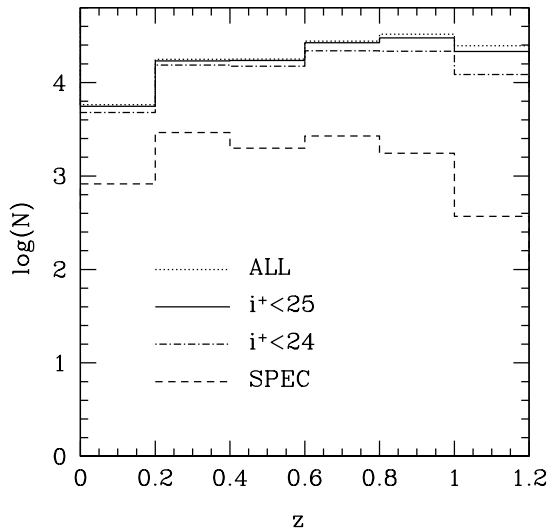}
\caption{The redshift distribution of the COSMOS sample. Highlighted are the distributions of sources with available good quality spectroscopy (dashed histogram), with $i^+<24$ (dotted-dashed histogram) and ${i^+}<25$ (solid histogram), while the overall population is denoted by the dotted histogram. Note the logarithmic scale on the y-axis.\label{fig:zdist}}
\end{figure}

In case the only redshift measurement is photometric, it is only adopted if the source has $i^+<25$ (as suggested by I09) and a good quality ($(z_{68\%}^{\rm up}-z_{68\%}^{\rm low})/(1+z_{\rm phot})<0.4$, with unflagged optical wavebands). The incompleteness due to these quality constraints is accounted for while computing source densities (Section~\ref{sec:dlfs}). This correction is given in Figure~\ref{fig:zincomp} for each of the IRAC bands used for sample selection (Section~\ref{sec:sampsel}). The figure shows the variation of the fraction of sources with $(z_{68\%}^{\rm up}-z_{68\%}^{\rm low})/(1+z_{\rm phot})<0.4$ and $i^+<25$ depending on observed magnitude in each of the IRAC-channels. We do not consider sources fainter than the magnitude beyond which the redshift completeness is smaller than 0.7 (Figure~\ref{fig:magcut}). The redshift distribution is shown in Figure~\ref{fig:zdist}, highlighting the fraction of the sources with available $z_{\rm spec}$, with $i^+<24$, with $i^+<25$, and the total population.

\begin{figure}
\plotone{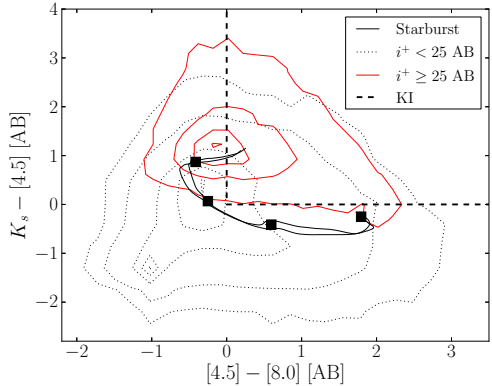}
\caption{The $K_s-[4.5]$ versus $[4.5]-[8.0]$ plot used to further characterise the $i^+>25$ population. The red solid contour shows the distribution of sources with $i^+\geq25$ (isocontours at densities of $\rm{N}=10,~100,~200,~265$), while the black dotted contour shows the distribution of sources with $i^+<25$ (isocontours at $\rm{N}=10,~100,~1000,~2000,~2800$). The black solid tracks, refer to the colours observed for a star-forming SED \citep[Arp220 and M82,][]{Polletta07} when redshifted between $z=0$ and $z=3$. Black squares mark $z=0,~0.5,~1,~2$, while the unmarked end refers to $z=3$. The dashed line delimits the selection region by the AGN KI-selection (Section~\ref{sec:gal}). We refer the reader to \citet{Messias12} for a more detailed discussion on this color-color region.\label{fig:kc24}}
\end{figure}

We note that by including only sources with $i^+<25$ in the study, highly obscured sources may be missed. Nevertheless, Figure~\ref{fig:kc24} shows that the $i^+>25$ population is composed mainly of starburst galaxies at $z\gtrsim1$ (with the bulk at $z\gtrsim2$) or AGN at any redshift. While this work does not address the $z\gtrsim1$ regime (Table~\ref{tab:zrange}), extremely obscured AGN at $z\lesssim1$ with $i^+\geq25$ will be missed. At this stage, one can only estimate an upper limit of the $i^+\geq25$ AGN by considering all sources inside the KI region to be AGN (they can also be $z\gtrsim2.5$ starbursts). This yields a 32\% rejection of the KI population when applying the $i^+<25$ cut. As the redshift of the faint KI-selected sources is unknown, we add in quadrature a 0.32 factor to the error budget while estimating AGN source densities (see below). This is a conservative value given not all KI sources with $i^+\geq25$ are AGN at $z\lesssim1$.

When computing measurement-associated errors, we consider in quadrature Poisson and $z_{\rm phot}$-induced errors ($\sigma_{\rm poi}$ and $\sigma_{z\rm{p}}$, respectively), and a factor ($f_{\rm o}$) to account for photometric redshift outliers (we adopt a conservative value of $f_{\rm o}=0.2$, I09 and Appendix~\ref{sec:zphagn}). In the case of the AGN sample, an additional factor ($f_{\rm A}$), accounting for the AGN with $i^+\geq25$, ought to be considered. For instance, the error associated to source density for the AGN sample is: $\sigma_{\rho}=\sqrt{\sigma_{\rm poi}^2+\sigma_{z_{\rm phot}}^2+(f_{\rm o}^2+f_{\rm A}^2)\rho^2}$, where $\rho$ represents the source density.

\section{Aperture photometry correction} \label{sec:apcorr}

We note that the reference IRAC photometry used to select our sample was measured using an aperture of 1$.\!\!^{\prime\prime}$9 and then scaled to total flux using a correction factor assuming a point source. However, such assumption is not always verified especially at low redshifts. At such distances, the physical size traced by a 1$.\!\!^{\prime\prime}$9 aperture is small ($<6\,$kpc, Table~\ref{tab:zrange}), thus probing only the central region of a galaxy.

\begin{figure*}
\epsscale{1.2}
\plotone{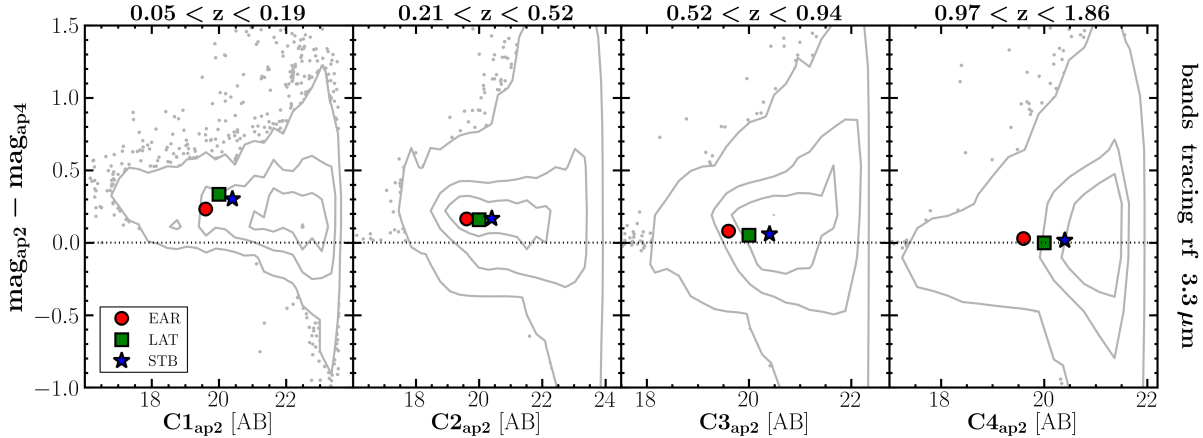}
\caption{Correcting the aperture photometry to ``total'' flux. This is particularly relevant in the lowest redshift intervals. The total-magnitude as estimated via a 1$.\!\!^{\prime\prime}$9 aperture is denoted as \emph{ap2}, while that estimated via a 4$.\!\!^{\prime\prime}$1 aperture is denoted as \emph{ap4}. The panels show the difference between \emph{ap2} and \emph{ap4} (y-axis) versus \emph{ap2} (x-axis). Each panel refers to different redshift bins where the rest-frame 3.3\,$\mu$m is traced, and, as a result, to different IRAC-filters: 3.6\,$\mu$m at $0.05<z<0.19$ (left-most); 4.5\,$\mu$m at $0.21<z<0.52$ (middle-left); 5.8\,$\mu$m at $0.52<z<0.94$ (middle-right); and 8.0\,$\mu$m at $0.97<z<1.86$ (right-most). The contours show the overall population distribution (the first isoline level refers to $\rm{N}=10$, while subsequent ones depend on sample maximum number density). The three data points (with arbitrary x-positions) show the average value $\rm{mag_{ap2}-mag_{ap4}}$ for each population (early-type as red circle, late-type as green square, and starburst as blue star) considering only sources with $\rm{mag_{ap2}<21\,AB}$. The y-axis ranges were fixed to allow for better comparison between redshift intervals. The horizontal dotted-line marks the level when \emph{ap2} agrees with \emph{ap4}.\label{fig:photcorr}}
\end{figure*}

Hence, we apply a statistical correction to the four IRAC bands depending on redshift and spectral-type (only for non-AGN populations) by comparing with the 4$.\!\!^{\prime\prime}$1 aperture (ap4) photometry. Figure~\ref{fig:photcorr} shows the magnitude difference between the two apertures versus the 1$.\!\!^{\prime\prime}$9 aperture (ap2) magnitude. Each panel refers to different redshift bins where the rest-frame 3.3\,$\mu$m is traced by the IRAC bands. Note, however, that the corrections are applied to all IRAC bands at each redshift bin and are different between bands. The discordance between the two apertures results from three factors: (\emph{i}) flux being missed by the ap2 photometry (shifting the whole distribution up); (\emph{ii}) contamination by neighboring sources (producing the scatter to positive y-axis values); (\emph{iii}) and bright nearby sources affecting the background measurement, which produces a flux underestimate at large apertures (producing the scatter to negative y-axis values). The correction aims at compensating for missing flux (\emph{i}), while being independent of artifacts (\emph{ii} and \emph{iii}). Only sources with $\rm{mag_{ap2}}<21\,$AB are considered, so that the correction value is unaffected by the large scatter at fainter magnitudes. It is clear that the lower the redshift, the larger is the correction.

We finally refer that the AGN sample is not corrected for this effect, as the emission from AGN sources is assumed to come from the nuclear region, thus not being affected by aperture photometry. Star-formation in the host-galaxy could in principle still produce such bias. However, we do not observe strong evidence for such a scenario in our IR-selected AGN sample (see Section~\ref{sec:d2t}).

\section{Estimating hot-dust and PAH luminosities} \label{sec:estimdust}

\begin{figure*}
\plottwo{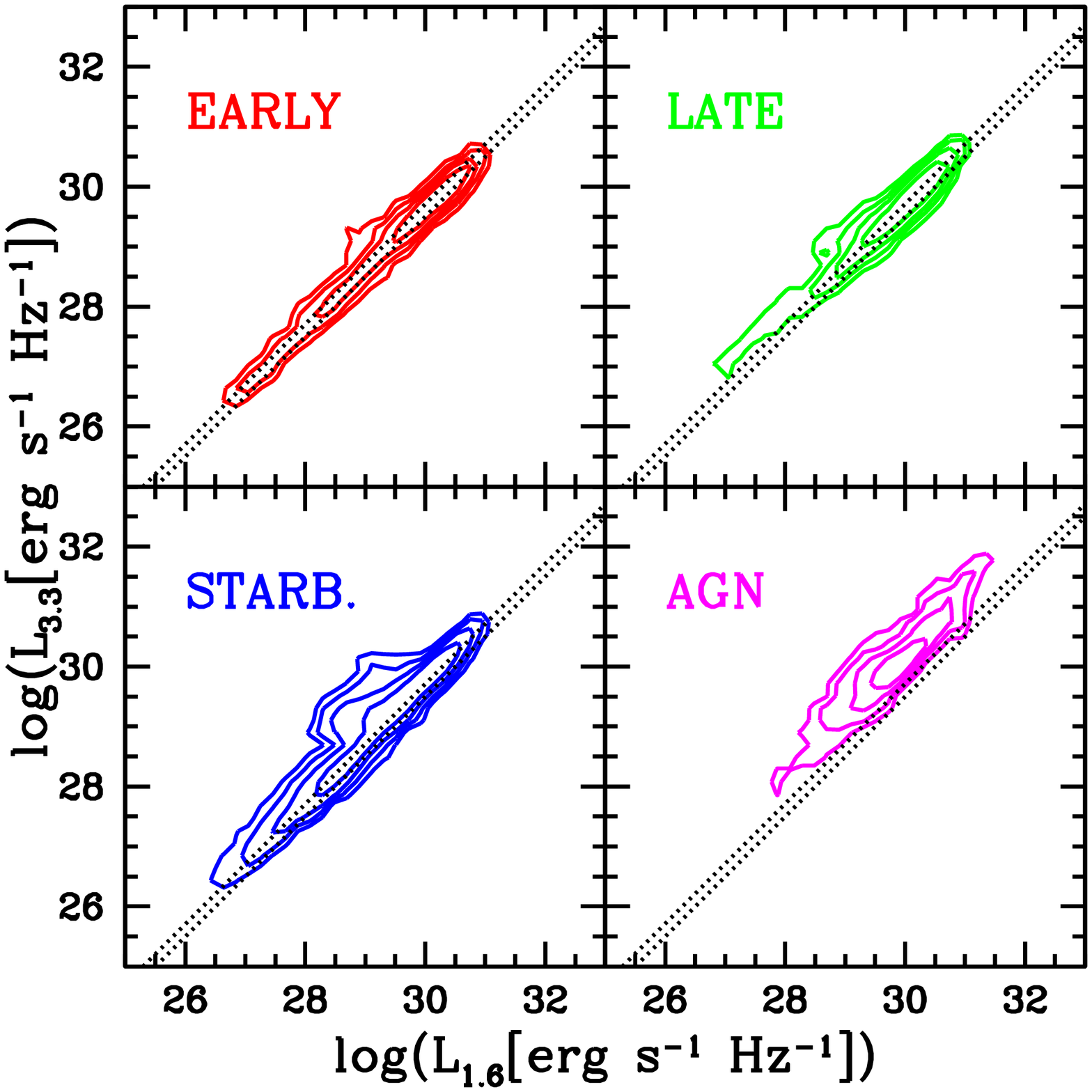}{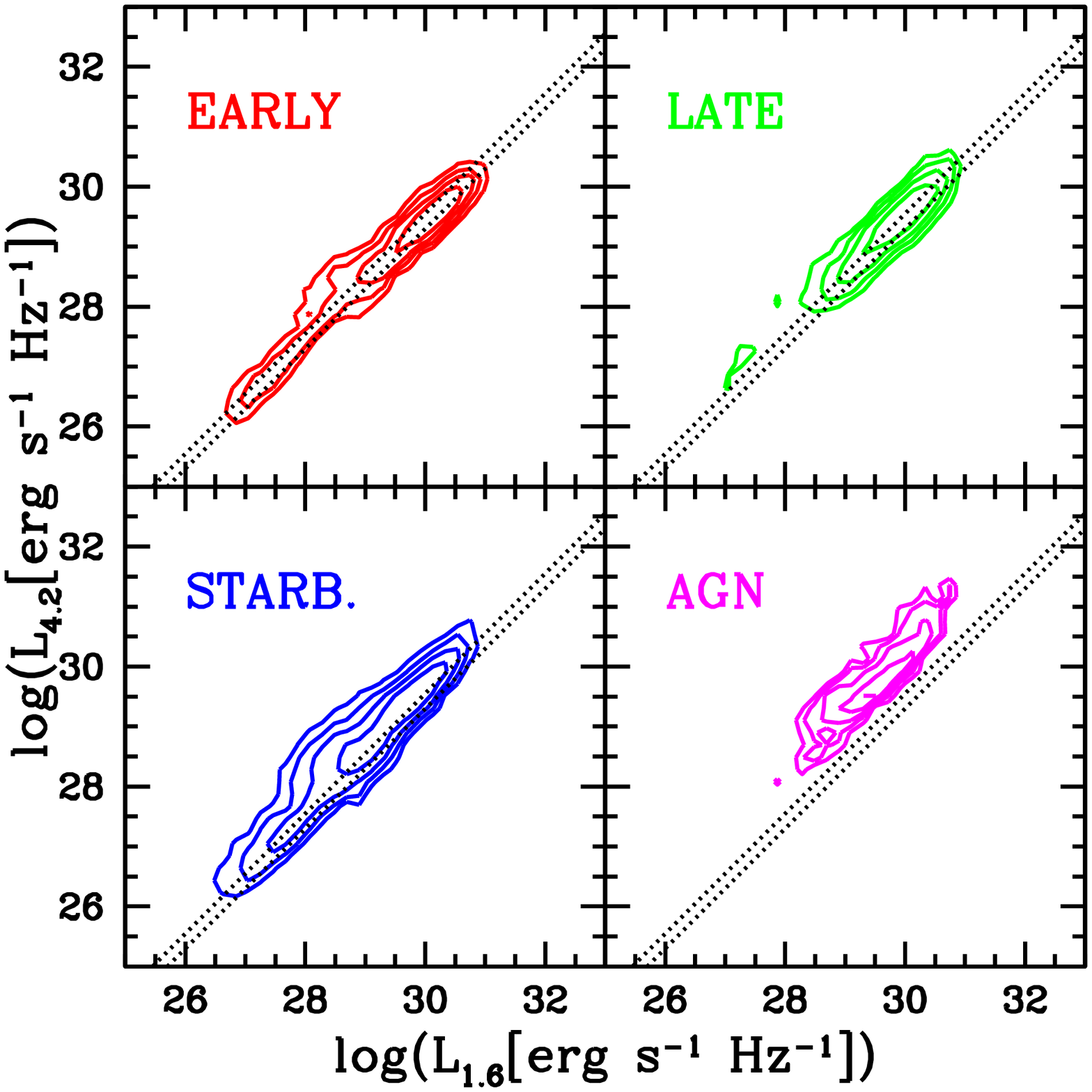}
\caption{Luminosities at rest-frames 1.6\,$\mu$m (L$_{1.6}$) and 3.3\,$\mu$m (L$_{3.3}$, left hand-side plot) or 4.2\,$\mu$m (L$_{4.2}$, right hand-side plot). Each panel refers to a different population: early-type (top left), late-type (top right), starburst (bottom left), and AGN host (bottom right). The contours are simply demonstrative of the sample distribution number-density and are defined based on the maximum source density in each plot, hence the isocontour levels differ between panels. The dotted lines delimit the region where pure stellar emission should fall. Any deviation from this region, clear in starburst and AGN populations, is assigned to non-stellar emission (hot-dust and PAH).\label{fig:cont33}}
\end{figure*}

In Section~\ref{sec:rfwl} we have mentioned the two rest-frame bands where hot-dust and PAH emission is expected: 3.3 and 4.2\,$\mu$m. However, at these wavelengths, one also needs to take into account stellar emission. Figure~\ref{fig:cont33} shows the estimated 1.6\,$\mu$m luminosities versus the observed --- hence including both stellar plus non-stellar emission --- 3.3\,$\mu$m (left hand-side plot) and 4.2\,$\mu$m (right hand-side plot) luminosities for each galaxy population considered in this study. The observe fluxes include both stellar and non-stellar emission. The regions between the dotted lines represent the locus where the SEDs dominated by stellar emission alone are expected to fall (see below). The elliptical data ``cloud'' tends to fall in this ``stellar region''. Any deviation from this region, clear in starburst and AGN populations, is assigned to non-stellar emission (hot-dust and PAH). This is the emission excess we aim to extract.

In order to disentangle stellar from non-stellar emission, we first estimate the stellar emission at 3.3 and 4.2\,$\mu$m for each galaxy. This is done in two steps. First we consider a reference wavelength where the observed flux is expected to come entirely from stellar emission. Using this, we then estimate the stellar emission at 3.3 and 4.2\,$\mu$m. The remaining flux is thus assumed to be due to contribution from non-stellar emission alone.

\begin{figure}
\plotone{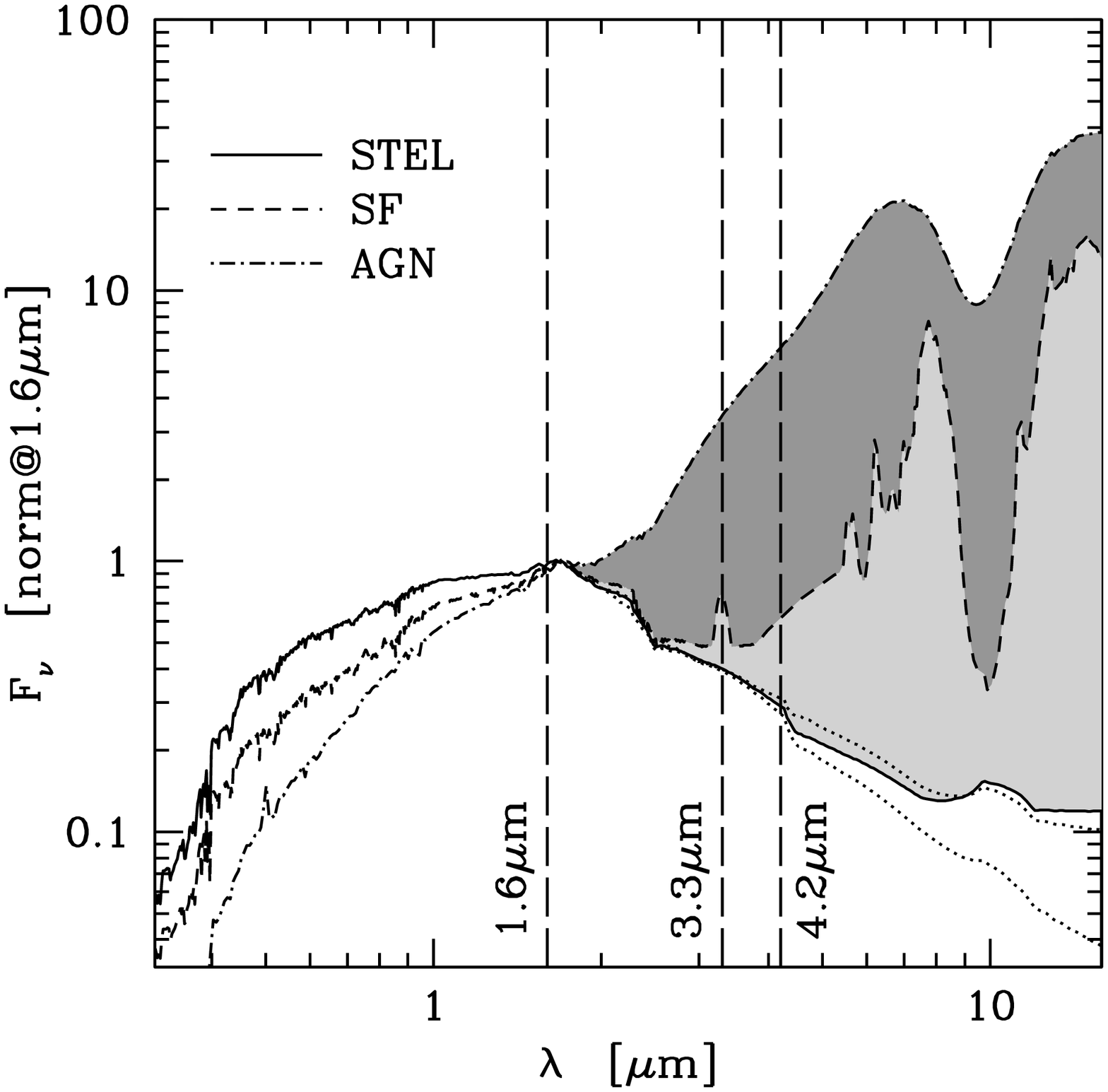}
\caption{Separating the IR emission into stellar and dust$+$PAH contributions. The solid line shows a 2\,Gyr old elliptical template (only stellar light) used for the conversions from rest-frame 1.6\,$\mu$m luminosities to 3.3\,$\mu$m and 4.2\,$\mu$m stellar luminosities (see text). The dotted lines show the range in flux (normalized to 1.6\,$\mu$m) between a 13\,Gyr old elliptical (lower-limit) and a 50\,Myr starburst (upper-limit). Together with the solid line, the dashed and dotted-dashed lines delimit, respectively, the dust contribution (shaded region) to the IR SED of Arp220 (a dusty starburst) and IRAS 19254-7245$_{\rm south}$ (an AGN host) galaxies \citep[templates from][]{Polletta07,Ilbert09}.\label{fig:ellmod}}
\end{figure}

The reference wavelength used to estimate the stellar emission is the stellar bump at rest-frame 1.6\,$\mu$m ($H$-band). This emission bump is frequently observed in galaxy SEDs (Figure~\ref{fig:ellmod}), being indicative of pure stellar contribution, as at $\lesssim2\,\mu$m no significant dust emission is expected. This is because UV/optical (stellar or AGN) emission is not enough to heat dust at the level for its cooling emission to dominate at such short wavelengths, or because any dust particle or PAH molecule in a strong radiation field is dissociated. Hence, at wavelengths below $\sim2\,\mu$m, only emission from the Wien tail of the spectrum due to the hottest dust grains (e.g., around Thermally Pulsating AGB, TP-AGB, stars or in a dust torus around an AGN) and from scattered AGN light is expected. Such emission can still be significant enough at 1.6\,$\mu$m to induce a systematic overestimate of the stellar emission, and consequently underestimate the non-stellar contribution. At this stage, we avoid quantifying such bias either for individual galaxies or for the statistically large sample used here.

\begin{figure}
\plotone{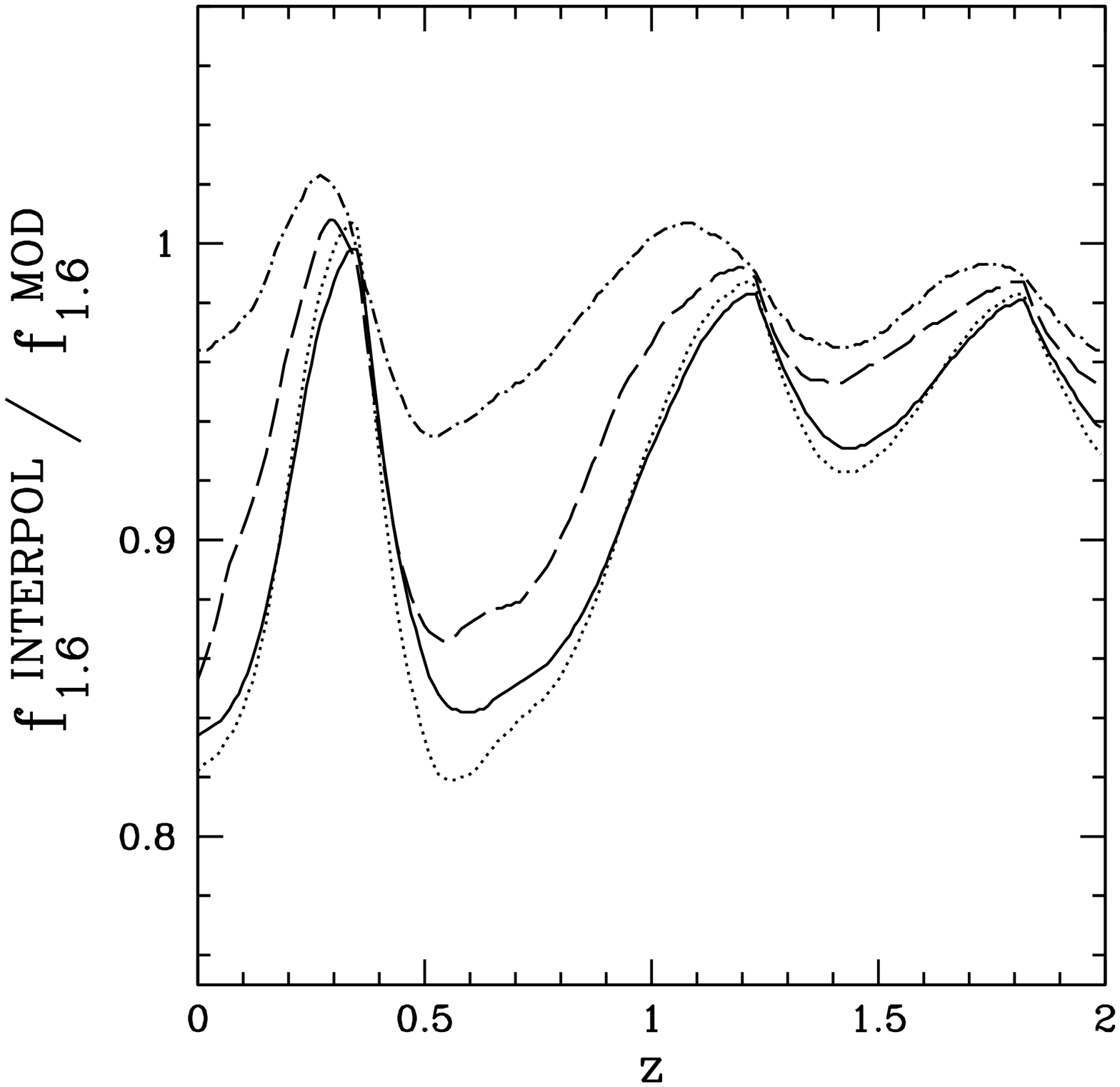}
\caption{The redshift-dependent effect of interpolating the galaxy SED in order to estimate the flux at rest-frame 1.6\,$\mu$m. The y-axis shows the ratio between the interpolated flux ($\rm{f^{INTERPOL}_{1.6}}$) and the actual model value ($\rm{f^{MOD}_{1.6}}$) at $1.6\,\mu$m. Different types are shown: early (solid line, 2\,Gyr Elliptical), late (dotted line, Sb), blue starburst (dashed line), and AGN (dash-dotted line, Mrk231). Except for the blue starburst (from S11), the adopted galaxy templates come from \citet{Polletta07}.\label{fig:interpol}}
\end{figure}

The flux at rest-frame 1.6\,$\mu$m is obtained through interpolation between the two wavebands which straddle this rest-frame wavelength at the source's redshift. However, interpolation is likely to underestimate the true rest-frame 1.6\,$\mu$m flux value depending on the source redshift and SED shape. This is evident from Figure~\ref{fig:interpol} where discrepancies between estimated and true value (always below the 20\% level) are shown for typical early (red) and late (green) galaxies, blue starbursts (blue), and AGN hosts (magenta). These trends were used to correct the interpolated 1.6\,$\mu$m flux for each galaxy at its respective redshift. Due to the absence of the $H$-band from the I09 catalogue, at $z\sim0$ the ratio in Figure~\ref{fig:interpol} is $<1$, because $J$ and $K_s$ passbands were used to interpolate the 1.6\,$\mu$m flux. The final corrected values are the ones shown in Figure~\ref{fig:cont33}.

With the estimated stellar flux at 1.6\,$\mu$m, the corresponding stellar contribution at 3.3 and 4.2\,$\mu$m is obtained. This is done with a pure stellar model (solid line in Figure~\ref{fig:ellmod}, corresponding to a 2\,Gyr old elliptical from \citealt{Polletta07}), the same used in Figure~\ref{fig:cont33} to compute the stellar-dominated locus. The conversion from 1.6\,$\mu$m stellar flux to that at 3.3 and 4.2\,$\mu$m is slightly redshift dependent, because the considered filters will probe slightly different rest-frame wavelength ranges. Hence, using the pure stellar model (Figure~\ref{fig:ellmod}), a conversion table was generated by convolving that stellar model with the NIR filters (from $J$-band to 8\,$\mu$m) at redshift steps of $\Delta{z}=0.01$.

We consider the underlying shape of the galaxy SED at 1.6--4.2\,$\mu$m, due to stellar emission alone, to be common to all galaxy populations referred in this study. This is a fair assumption for a universal initial mass function. The stellar emission in this spectral regime originates in cold stars, which live longer, thus producing a constant SED shape over a wide range of ages. Such assumption may be affected by strong differential obscuration affecting the rest-frames 1.6\,$\mu$m and 3.3 or 4.2\,$\mu$m. However, this will only occur in extremely obscured systems \citep[e.g.,][]{daCunha08}, which are rare. Figure~\ref{fig:ellmod} supports this assumption, showing how similar the 1.6--4.2\,$\mu$m stellar SED is between a 13\,Gyr old elliptical and a 0.05\,Gyr old starburst (models from I09).

The next step is to separate the hot-dust emission from the PAH emission at 3.3\,$\mu$m. In order to do this, we use the results from \citet{daCunha08} who find that a gray-body model of temperature 850\,K and with a dust emissivity index of 1 is optimal to fit galaxies' continuum emission in the 3--5\,$\mu$m spectral range. Assuming this continuum shape, we normalize it to the 4.2\,$\mu$m dust emission flux to estimate the flux at 3.3\,$\mu$m. Removing this from the total non-stellar 3.3\,$\mu$m emission, one estimates the PAH 3.3\,$\mu$m emission flux. This procedure is only applied to non-AGN sources, given that \citet{daCunha08} only considered non-AGN sources. In the case of AGN, a dominant power-law continuum shape is more likely. Hence, we estimate the 3.3\,$\mu$m continuum emission in AGN via interpolation of the two bands straddling the band tracing rest-frame 3.3\,$\mu$m, assuming $f_\nu\propto\nu^\alpha$.

Henceforth, whenever ``PAH emission'' is mentioned, we refer to the 3.3\,$\mu$m PAH emission flux estimated by the above procedure, while whenever ``hot-dust emission'' is mentioned, we refer to the estimated 4.2\,$\mu$m non-stellar emission flux assigned to hot-dust emission alone.

\section{The evolution of the Dust-to-total and PAH-to-total luminosity ratios} \label{sec:d2t}

\begin{figure*}
\plottwo{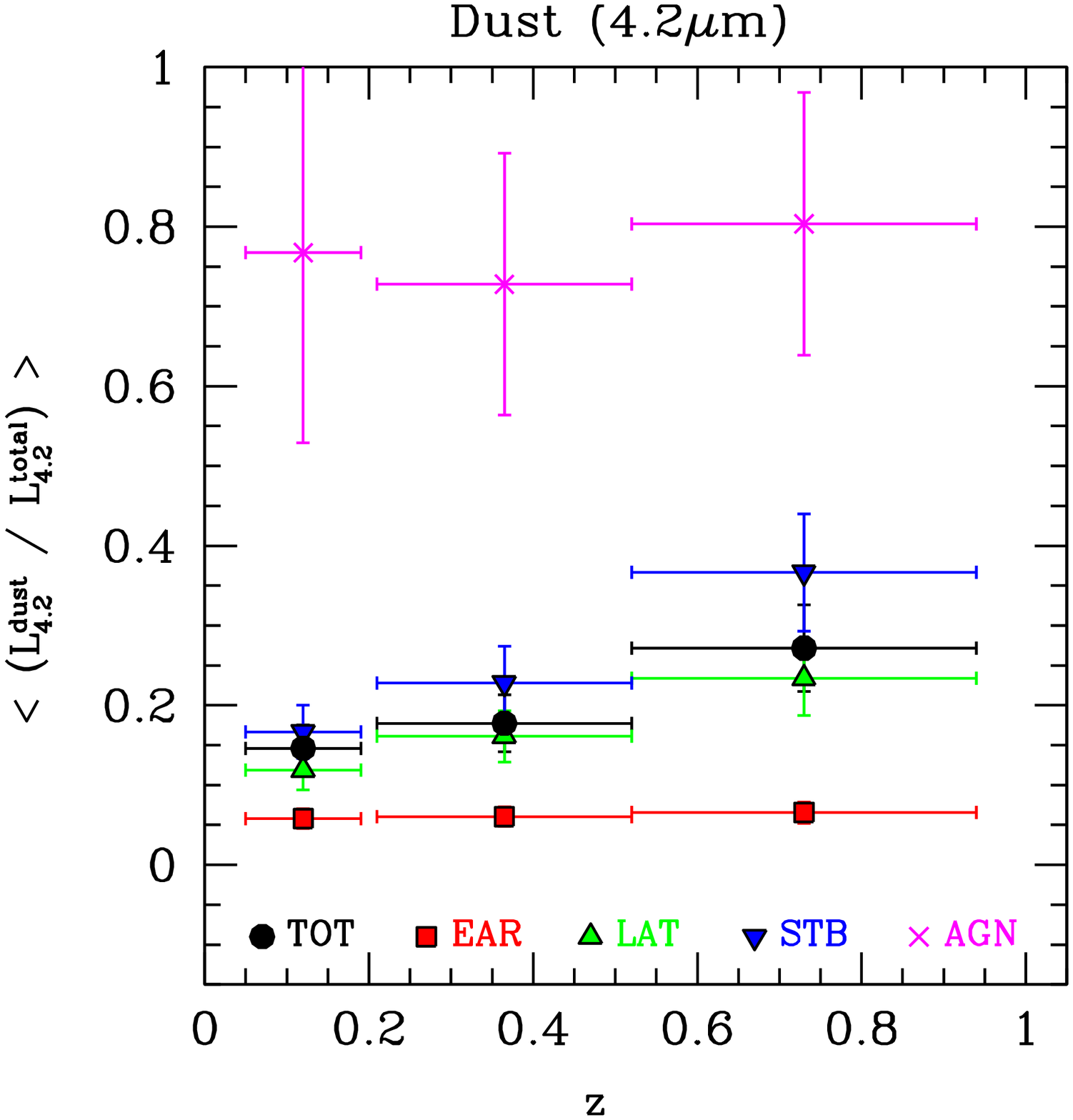}{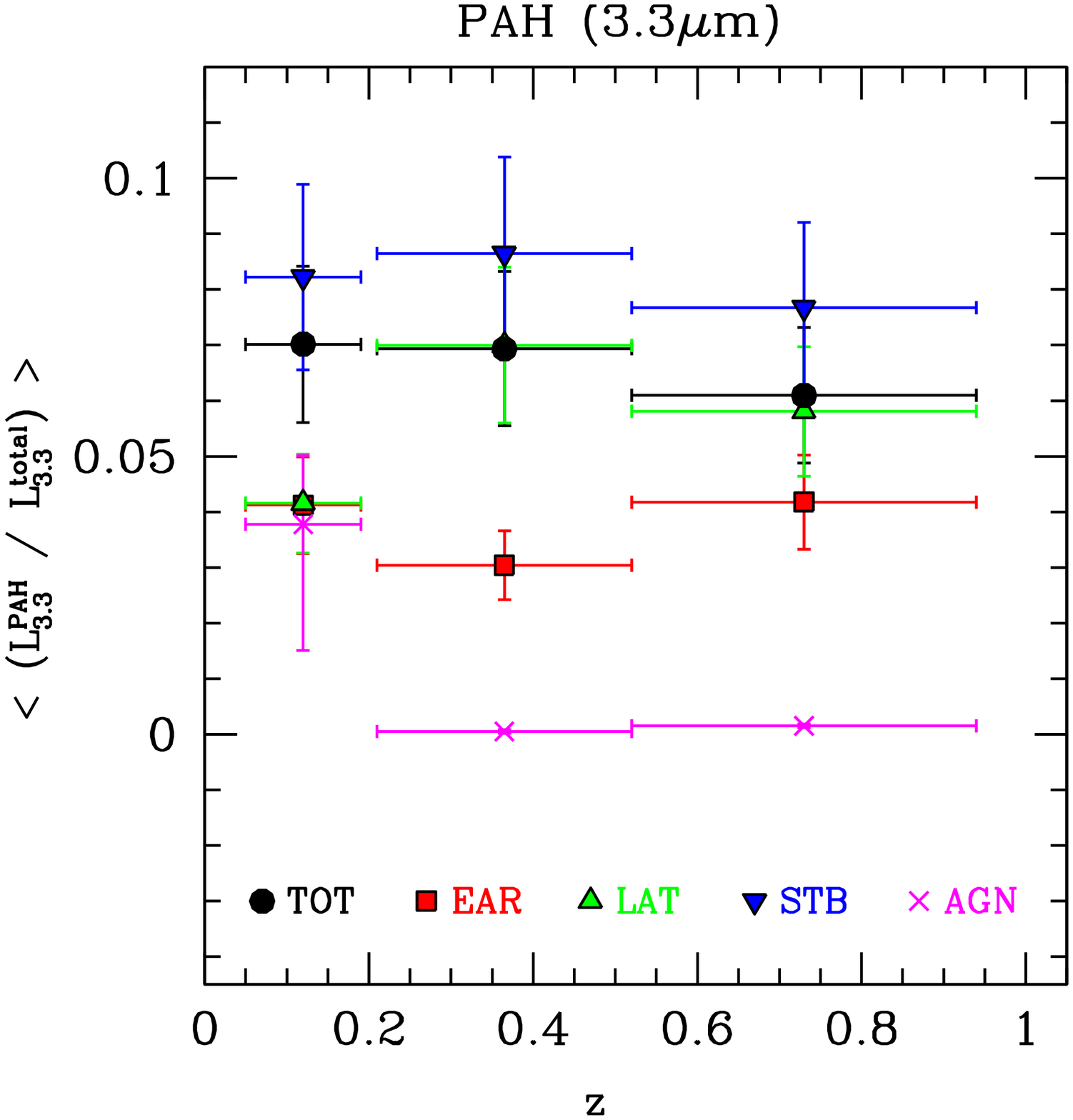}
\caption{The dust-to-total (at 4.2\,$\mu$m, left hand-side) and PAH-to-total (at 3.3\,$\mu$m, right hand-side) luminosity ratio dependency on redshift. Different populations are considered: total (black circles), early (red squares), late (green upward triangles), starburst (blue downward triangles), AGN (magenta crosses). Note the different y-axis scales. The values only take into account galaxies brighter than the magnitude beyond which 5\% of the sources do not have a 4.2$\,\mu$m flux estimate (Figure~\ref{fig:magcut})\label{fig:d2t}}
\end{figure*}

Figure~\ref{fig:d2t} presents the evolution of the average dust-to-total (left hand-side) and PAH-to-total (right hand-side) luminosity ratios. It is clear that the average contribution of hot-dust at 4.2\,$\mu$m IR-selected AGN remains constant with redshift around the 80\% level. This constancy results from using IR AGN selection criteria alone. The goal here is to independently trace different dust-heating mechanisms (AGN activity versus star-formation). Note also the extremely low PAH contribution in AGN, consistent with PAH-annihilation by the AGN radiation or emission-dilution in the bright AGN-dominated SED. The increase at the lowest redshift interval for the AGN population may be explained by the aperture photometry adopted to estimate colors. With decreasing redshift, a fixed aperture will probe more nuclear regions, enabling the detection of lower luminosity AGN. Being less powerful, these AGN allow for the existence of PAH molecules as opposed to more luminous AGN, the dominant population with increasing redshifts, which shows average PAH-to-total luminosity ratios close to zero.

While the dust-to-total luminosity ratio for early-type galaxies is consistent within $\sim1\sigma$ to be constant ($\sim5\%$) with redshift, it increases with increasing redshift for late-type and starburst galaxies. The starburst and early-type populations present a constant PAH-to-total luminosity ratio (within the errors) with redshift, while the late-type population shows a decrease at the lowest redshift interval.

The PAH-to-dust luminosity ratio seems to vary slightly. For instance, the ratio for the starburst sample is $0.49\pm0.16$ at $0.05<z<0.19$, $0.40\pm0.14$ at $0.21<z<0.52$, and $0.21\pm0.10$ $0.52<z<0.94$. In the literature, the 3.3-PAH feature strength is compared to colder-dust (emitting at 8--1000\,$\mu$m) instead, and there is no evidence for evolution: \citet{Dasyra09} estimate for a sample at $\overline{z}=0.87$ an average 3.3-PAH to 8--1000\,$\mu$m luminosity ratio of $\sim10^{-3}$, while \citet{Siana09} find the ratio to be $8.5\pm4.7\times10^{-4}$ in a $z=3.074$ source \citep[see also][]{Magnelli08,Sajina09}, being both values in agreement with those found for lower-redshift starburst galaxies \citep{Imanishi10,Yamada13}.

\begin{figure*}
\plottwo{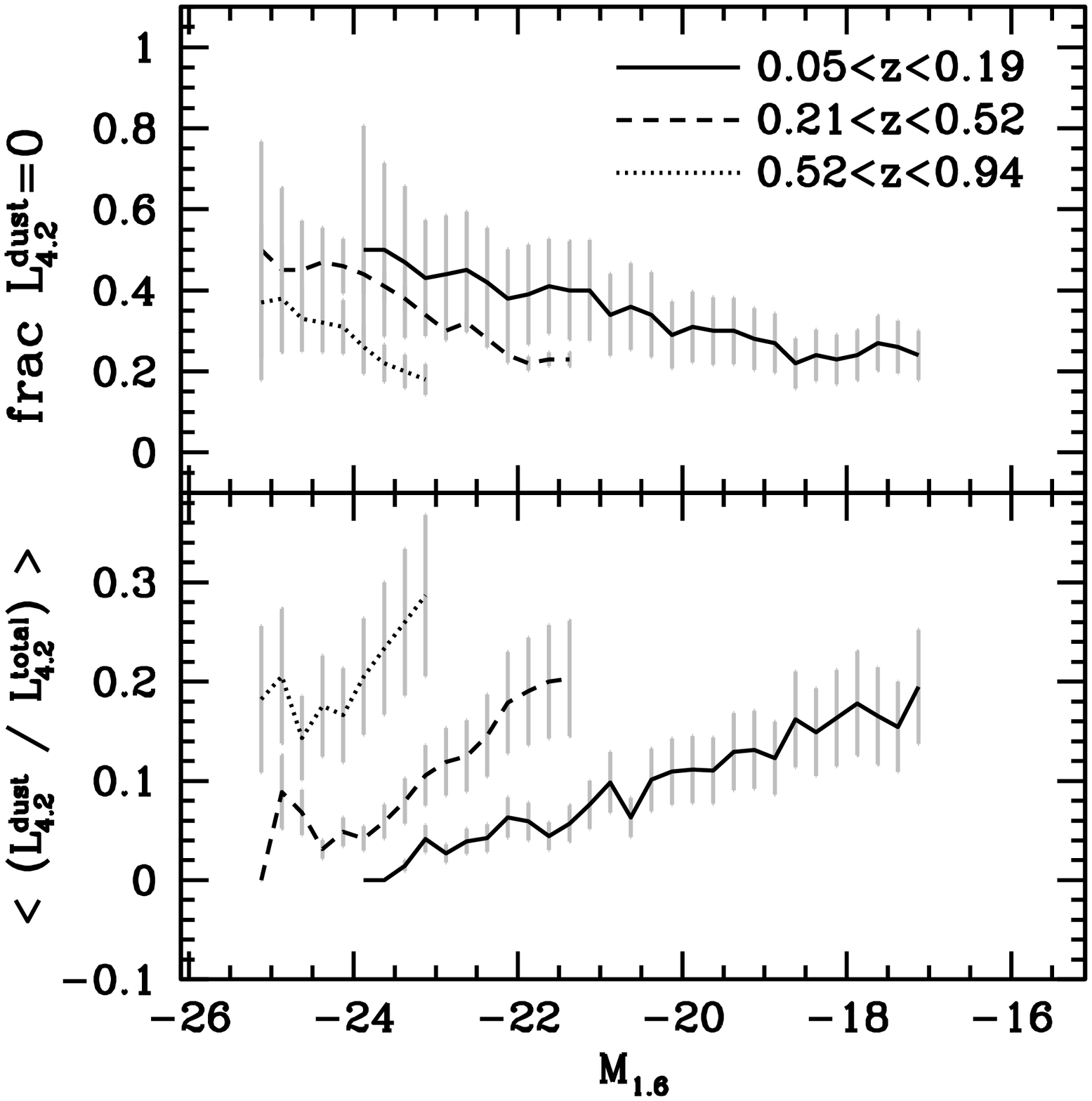}{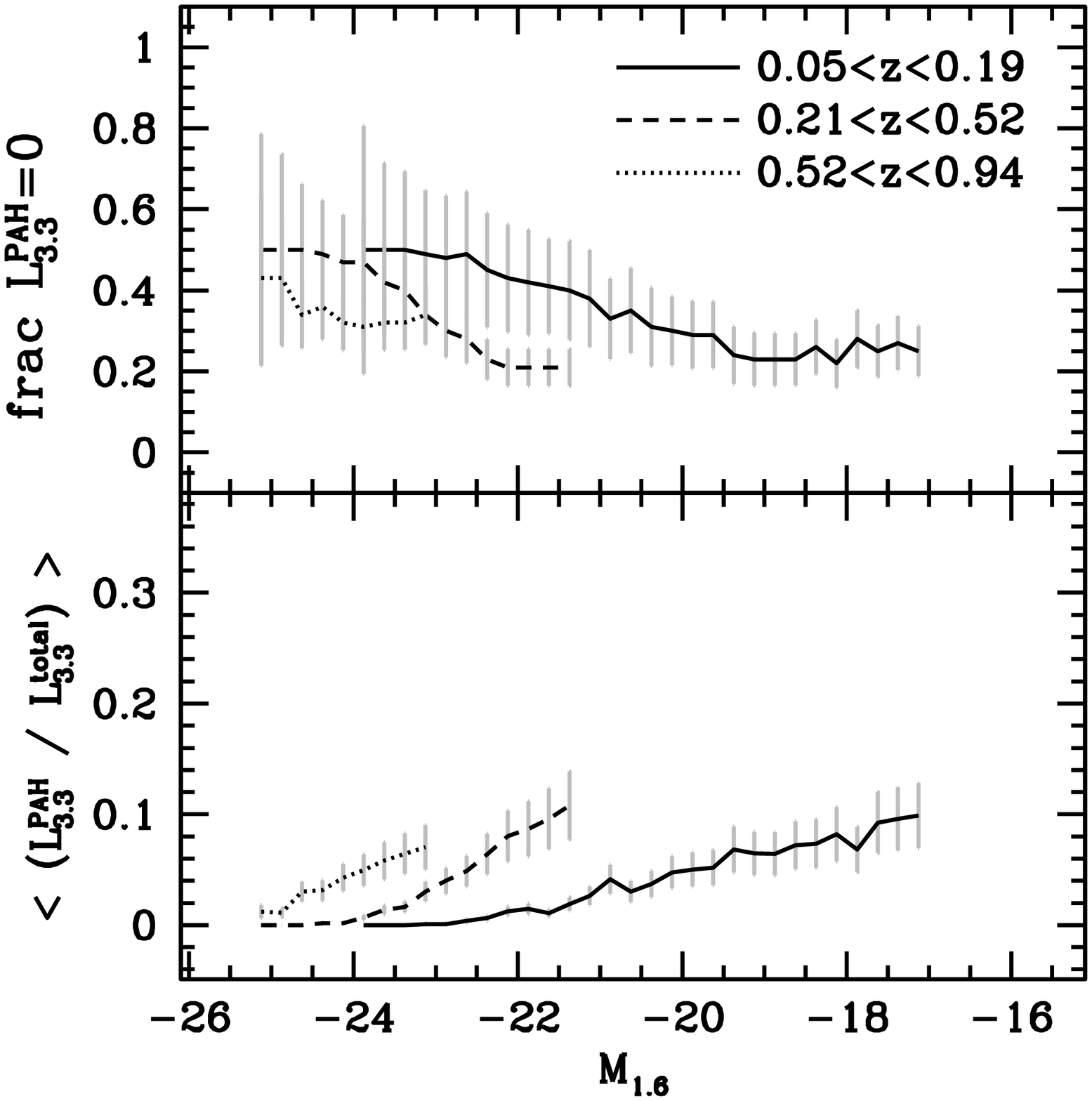}
\caption{The dust-to-total (at 4.2\,$\mu$m, left hand-side) and PAH-to-total (at 3.3\,$\mu$m, right hand-side) luminosity ratio dependency on 1.6\,$\mu$m luminosity and redshift (bottom panels): $0.05<z<0.19$ as solid line, $0.21<z<0.52$ as short dashed line, and $0.52<z<0.94$ as dotted line. Top panels show the dependency on 1.6\,$\mu$m luminosity and redshift of the fraction of sources which show no dust (left) or PAH (right) emission. The trends were trimmed according to the magnitude beyond which 5\% of the sources do not have a 4.2$\,\mu$m flux estimate (Figure~\ref{fig:magcut}).\label{fig:d2tlum}}
\end{figure*}

The fact that we find an increase of the average dust-to-total luminosity ratio with redshift does not necessarily mean the sample is progressively incomplete at higher redshifts (i.e., missing pure stellar emission sources, $L_{4.2}^{\rm dust}=0$). Figure~\ref{fig:d2tlum} (bottom panel in left hand-side plot) shows that higher redshift galaxies are indeed dustier in the luminosity range where all samples are complete. However, below the luminosity completeness cuts (Section~\ref{sec:sampsel}), one cannot assess the trends for the high-$z$ intervals. The decrease in the number of galaxies showing no dust emission ($L_{4.2}^{\rm dust}=0$) with decreasing 1.6\,$\mu$m luminosity is likely to be real (top panel in left hand-side plot), as the slope of the relation is constant up to the brightest luminosities.

\section{Dust Luminosity Density Functions} \label{sec:dlfs}

\begin{figure*}
\plotone{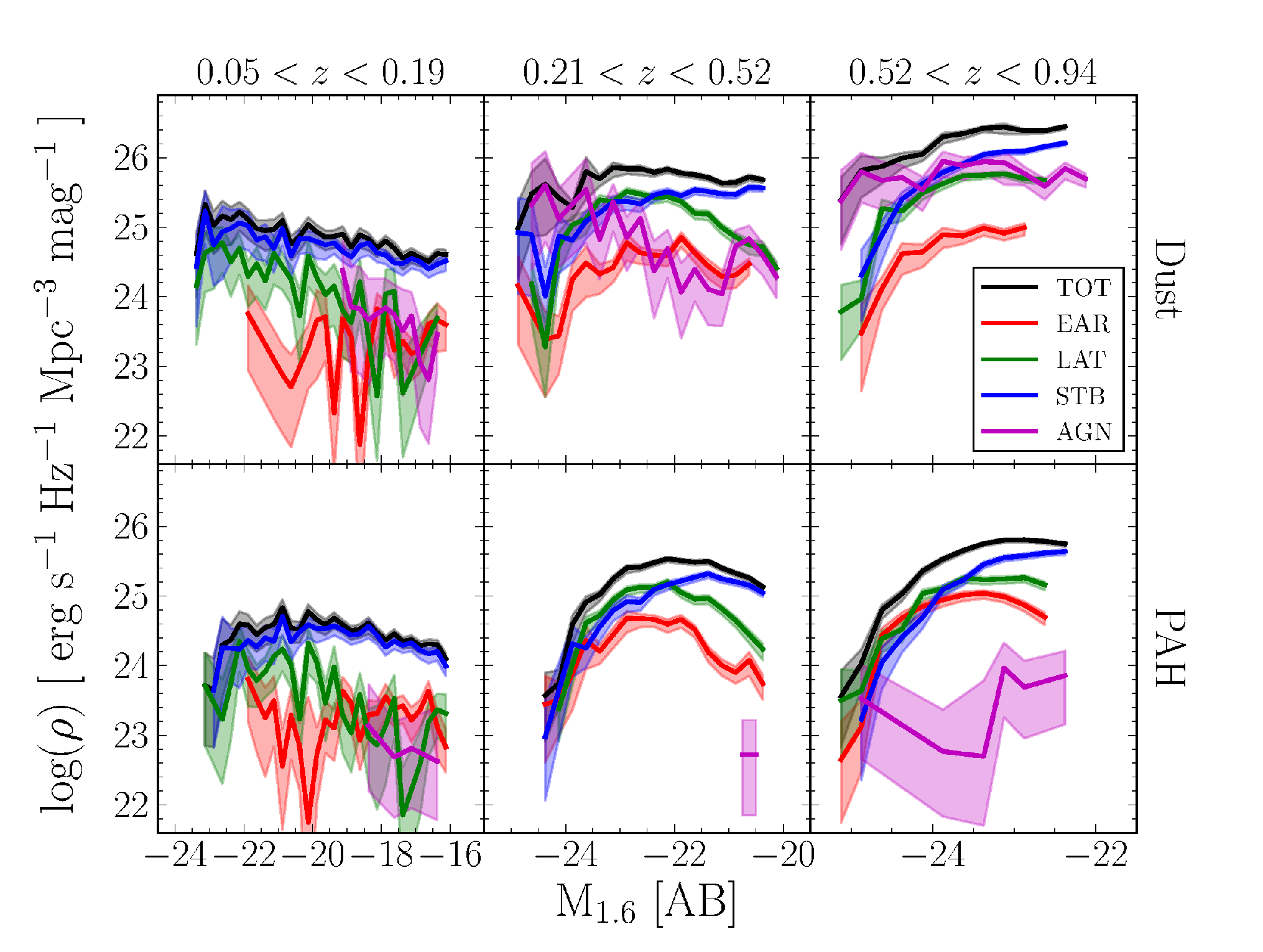}
\caption{Hot-dust (top rows) and PAH (bottom rows) luminosity densities as a function of $H$-band absolute magnitude (x-axis), redshift (each column refers to a different interval) and galaxy spectral type (color-coding as in Figure~\ref{fig:cont33}). For improved visualisation, different x-axis ranges are adopted. Intervals are not plotted whenever more than 30\% of the sources had no ${\rm V_{max}}$ estimate.\label{fig:lf33d}}
\end{figure*}

\begin{figure*}
\plottwo{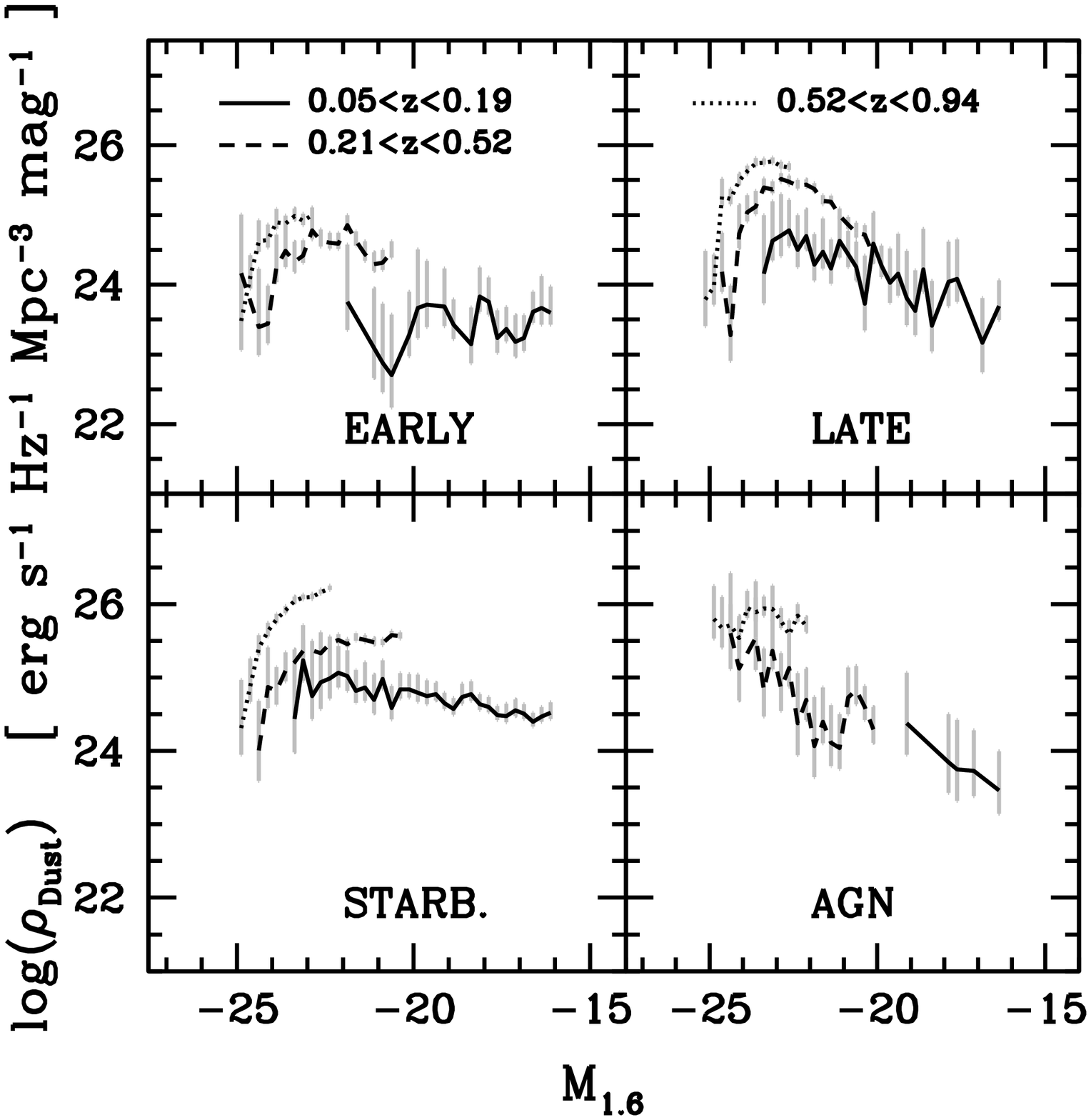}{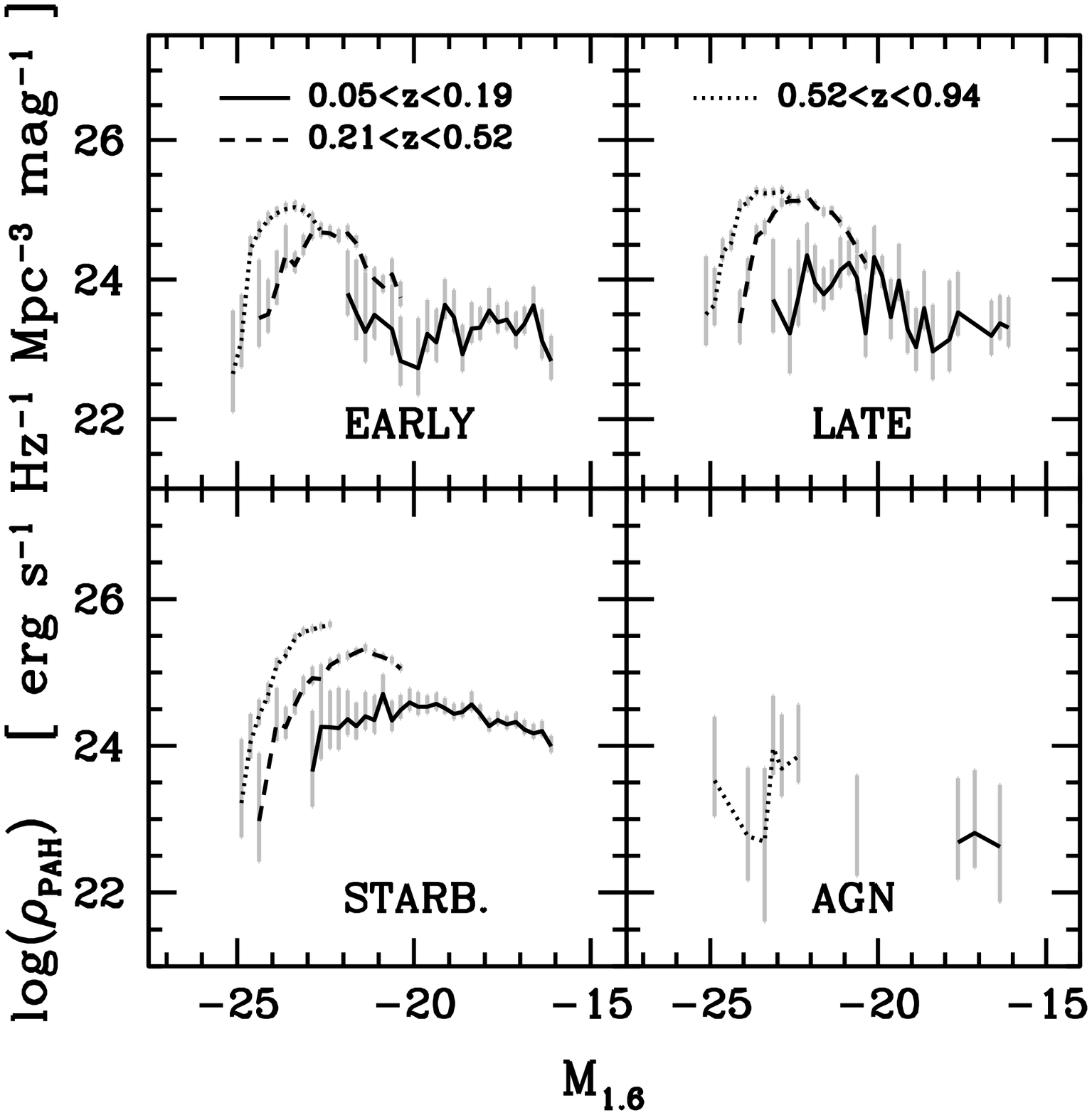}
\caption{Comparing the hot-dust (left hand-side plot) and PAH (right hand-side plot) LDFs for each galaxy population between redshift bins. Line coding as in Figure~\ref{fig:d2tlum}. Intervals are not plotted whenever more than 30\% of the sources had no ${\rm V_{max}}$ estimate.\label{fig:lf33dz}}
\end{figure*}

In this section we measure the hot dust and PAH luminosity density functions (LDFs) and their evolution with redshift for different populations of galaxies. Figure~\ref{fig:lf33d} compares the populations at different redshift intervals (shown in individual panels). The populations are color-coded: total (black), early-type (red), late-type (green), starburst (blue), and AGN hosts (magenta). Figure~\ref{fig:lf33dz} shows the evolution with redshift of the LDFs for each population (shown in individual panels). The volume associated with each galaxy is based on the flux limit of the sample and the $k$-correction, derived from the galaxy's own SED (as given by the observed multi-wavelength photometry), and obtained through the ${\rm 1/V_{max}}$ method \citep{Schmidt68}.

The dust LDFs enable us to evaluate how much dust is contributing to the IR radiation at any given luminosity and redshift for each galaxy population. Note that 1.6\,$\mu$m luminosities can be taken as a proxy to stellar mass assuming 1.6\,$\mu$m to be dominated by stellar emission. Therefore, one could assume here, as a proxy, the dust contribution as a function of galaxy stellar mass. Also, using a common luminosity discriminator, allows for a direct comparison between the dust and PAH LDFs.

Although AGN hosts are rare ($<4\%$ in our sample), this population clearly dominates the bright-end of the dust LDFs at least at $z>0.52$. The opposite is seen in the PAH LDFs, where the AGN population is the weakest contributor at all redshifts.

Figure~\ref{fig:lf33dz} shows evidence for a number-density evolution in dust LDFs for AGN. If one assumes the dust-to-total luminosity ratio to be constant in IR-selected AGN (Figure~\ref{fig:d2t}), then one would not expect to see any evolution with redshift at fixed luminosities in the dust luminosity density ($\rho_{\rm D}$) assuming a constant number-density. However, this seems to be the case, especially from $0.52<z<0.94$ to $0.21<z<0.52$. Further confirmation of this result may be accomplished by extending this study to deep IRAC coverages available, e.g., in the GOODS fields.

As shown in Figure~\ref{fig:lf33dz}, the major evolution in $\rho_{\rm D}$ and PAH luminosity density ($\rho_{\rm PAH}$) for the non-AGN populations mainly appears at the highest luminosities. That is, the most massive galaxies present the strongest evolution with redshift in both $\rho_{\rm D}$ and $\rho_{\rm PAH}$.

Regarding the PAH LDFs in Figure~\ref{fig:lf33d}, the faint-end is always dominated by the starburst population, followed by the late-type, and then the early-type galaxies. At high luminosities, the three non-AGN populations seem to contribute at comparable rates. This, however, does not happen in the dust LDFs, where the early-type population is the weakest contributor, while late-type and starburst populations have the highest contribution. This is explained in Figure~\ref{fig:d2t}, which shows that the higher the redshift, the larger is the difference in dust-to-total luminosity ratio between the early-type galaxies and late-type or starburst galaxies, while this is not the case for the PAH-to-total luminosity ratios.

The reason why this happens is not straightforward, but assuming the gray-body model scaling should instead more-or-less correlate to 3.3-PAH strength, one can conclude the following scenarios: (i) extra flux in early-type galaxies at 3.3\,$\mu$m unaccounted by PAH emission; (ii) or extra emission at 4.2\,$\mu$m in starburst and late-type galaxies. The former could imply higher gray-body temperatures ($>$850\,K) in early-types, for instance, induced by low-luminosity AGN, which were missed by the adopted AGN selection, while (ii) could be explained by \citet{Mentuch09}, who show evidence for extra emission in the 2--5\,$\mu$m spectral range unaccounted by the same gray-body model we assume plus PAH emission. They also find that this extra emission correlates with star-formation \citep{Mentuch09,Mentuch10}, thus explaining why we observe the enhancement in starburst and late-type galaxies and not in early-types. This reasoning also implies that the extra-emission heating mechanisms are in part different from those of the 3.3-PAH feature. One source of extra-emission at $\sim4.2$ could be the Br$_\alpha$ emission line (4.05\,$\mu$m). If present, it would imply an underestimate of the 3.3-PAH flux. However, the star-forming sample gathered by \citet{Yamada13} shows that the luminosity ratio between Br$_\alpha$ and 3.3-PAH feature is $\rm{Lum(Br_\alpha)/Lum(3.3)=0.08\pm0.04}$, which is not enough to account for the difference pointed out in Figure~\ref{fig:lf33d}.

\section{Evolution of dust luminosity density} \label{sec:ldz}

\begin{figure*}
\epsscale{1.15}
\plottwo{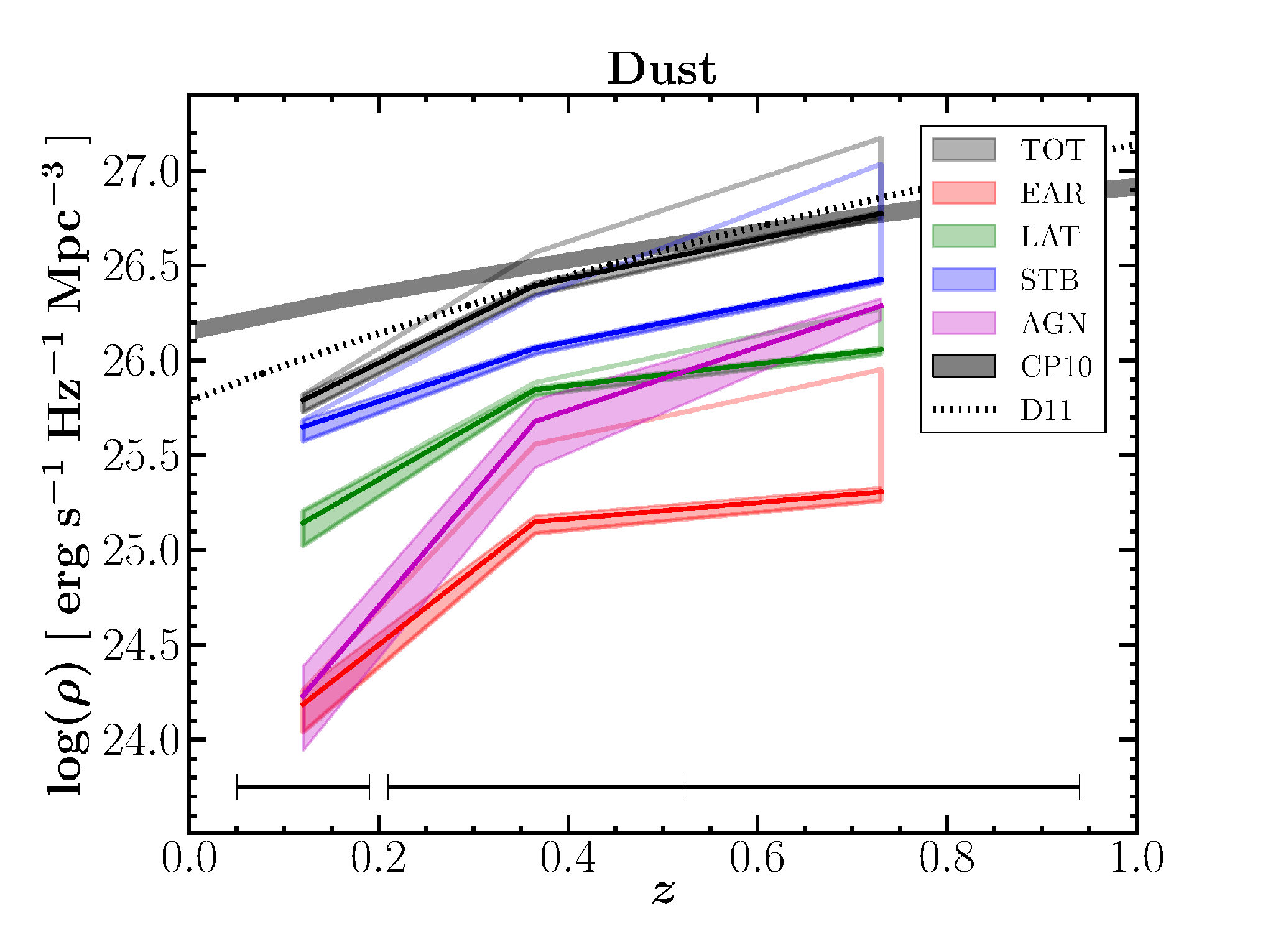}{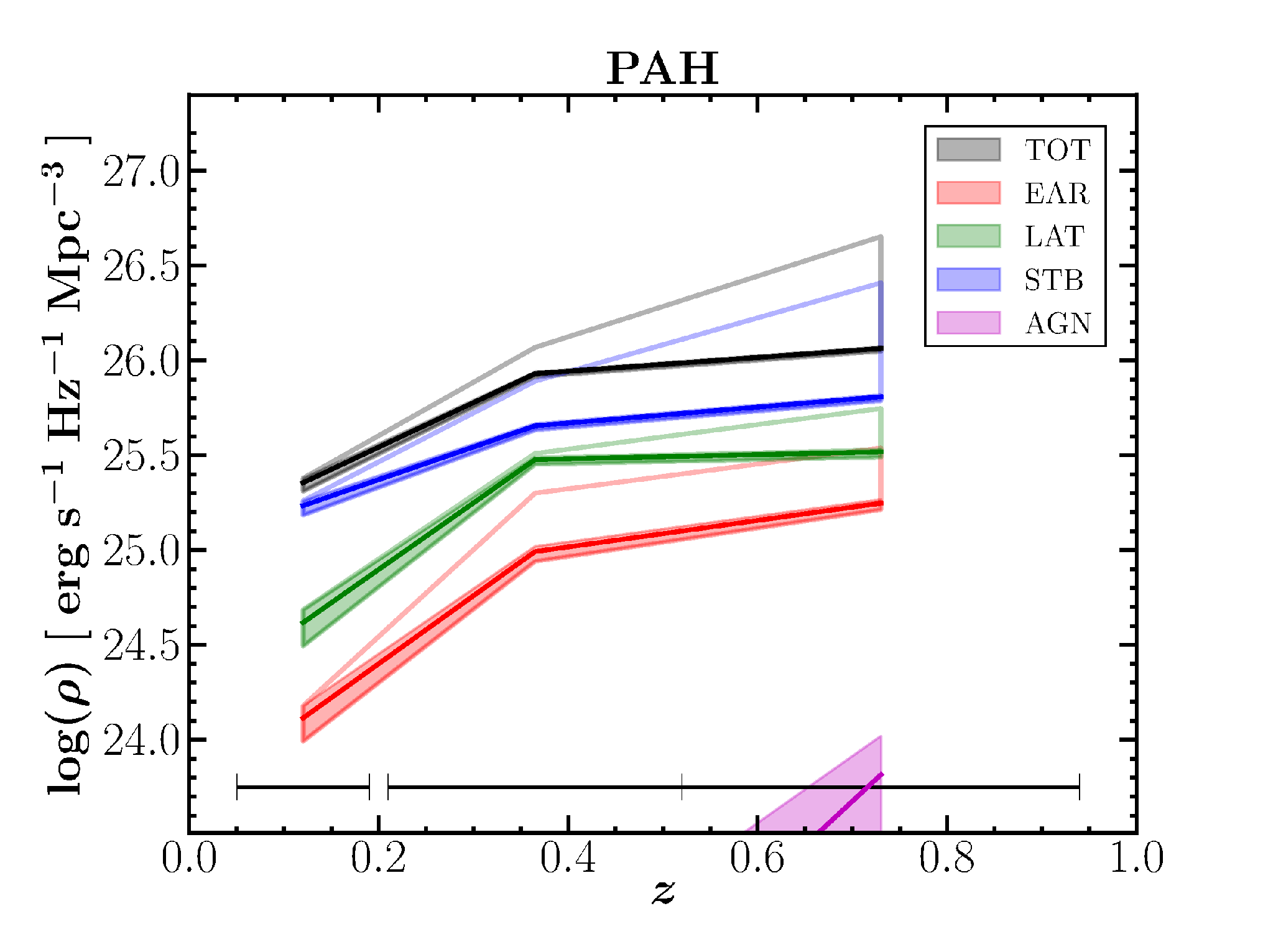}
\caption{Evolution with redshift of the dust and PAH luminosity densities (left and right plots, respectively) depending on galaxy type (color-coding as in Figure~\ref{fig:cont33}). Transparent regions show the associated $1\sigma$ error. Empty regions show the results if a correction for incompleteness is attempted (only for the total and non-AGN populations). The redshift intervals are indicated at the bottom as horizontal error bars. The gray shaded region in the left plot refers to the evolution trend of the cold-dust luminosity density alone derived from FIR observations \citep{Magnelli09,CharyPope10} and it is scaled to the $\rho_{\rm D}$ value of the total population at $0.52<z<0.94$. For improved visual comparison, the y-axis scale was matched between the two plots. However, this means $\rho_{\rm PAH}$ estimates for the AGN population fall mostly out of the plot.\label{fig:zlum}}
\end{figure*}

In Figure~\ref{fig:zlum}, we present the evolution of $\rho_{\rm D}$ and $\rho_{\rm PAH}$ since $z\sim$1 (left and right hand-side, respectively). We also compare $\rho_{\rm D}$ to the cold-dust luminosity density estimated from FIR observations \citep[70--500\,$\mu$m, dark gray shaded region,][]{Magnelli09,CharyPope10}, linearly translated to obscured star-formation. However, comparing the luminosity density at present time to that at $0.52<z<0.94$, the drop in $\rho_{\rm D}$ is more significant than that of the cold-dust luminosity density, by $\sim0.5\,$dex.

Such decrease is driven mostly by the non-starburst populations. While one may attribute part of the decay to the strong evolution shown by AGN hosts, whose FIR SED is related to star-formation and not to AGN activity, the late-type population also plays a role shaping such difference between hot and cold $\rho_{\rm D}$ evolution. However, even the starburst population shows a different rate of evolution compared to that of the far-IR evolution: the drop in $\rho_{\rm D}$ is still greater by $\sim0.3\,$dex compared to that of cold-dust.

\begin{figure}
\plotone{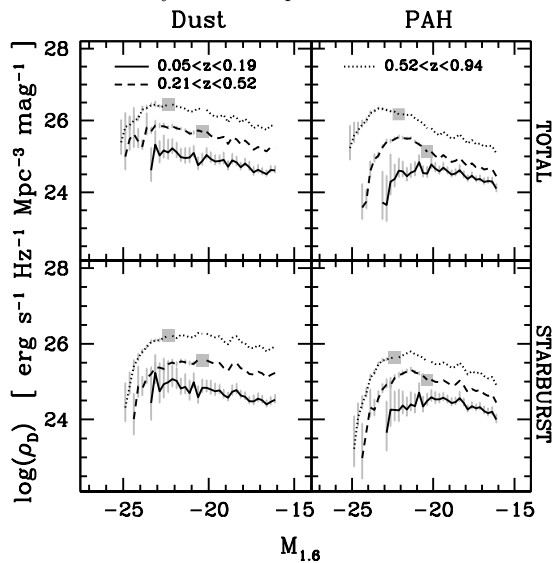}
\caption{Extending the dust (left hand-side panels) and PAH (right hand-side panels) LDFs below the completeness cut (indicated with gray square) assuming a similar shape as that observed for lower redshift intervals. Only LDFs for the total (upper panels) and starburst (lower panels) populations are displayed. Line coding as in Figure~\ref{fig:d2tlum}.\label{fig:comp}}
\end{figure}

We recall that the faint population, gradually missed with increasing redshifts by our completeness cuts, is likely a significant contributor to the overall $\rho_{\rm D}$ and $\rho_{\rm PAH}$ given the higher hot-dust and PAH contribution observed in their SEDs (Figure~\ref{fig:d2tlum}). This implies that the discrepancy between the hot and cold dust regimes is likely larger. Also, Figure~\ref{fig:lf33dz} shows a slower evolution in $\rho_{\rm D}$ and $\rho_{\rm PAH}$ at fainter luminosities. Accepting this as true, one can then consider the faint-end shape of the LDFs (i.e., that below the completeness cut) is unchanged from a redshift interval to a subsequent lower one and estimate a completeness correction. This is shown in Figure~\ref{fig:comp} for the total and starburst populations. Integrating a given LDF at luminosities fainter than the completeness cut will thus provide an estimate of the extra $\rho$ missed by our sample. Applying this correction yields extra luminosity density represented as the empty regions shown in Figure~\ref{fig:zlum} (only for the total and non-AGN populations). These corrections are expected to provide upper-limits for the $\rho$ values.

Recent work on the 3.3\,$\mu$m PAH feature with \emph{Spitzer} and \emph{Akari} out to $z\sim2$ does not show an evolution with redshift of the luminosity ratio $L_{3.3\mu\rm{m}}/L_{8-1000\mu\rm{m}}$ \citep{Magnelli08,Dasyra09,Sajina09,Siana09,Imanishi10,Yamada13}. However, comparing with hot-dust, our study seems to support slight variation of the hot-dust-to-PAH luminosity ratio with redshift.

\citet{Dunne11} find evidence for the stellar nurseries (birth clouds) to be producing a higher fraction of obscuration than the diffuse inter-stellar medium at high-redshift. Figure~\ref{fig:zlum} corroborates such finding when comparing our results with the cold-dust luminosity density evolution. This explains why high-redshift star-forming galaxies are more obscured \citep{Choi06,Villar08,Garn10,Dunne11}, as also pointed out by \citet{Dunne11}.

Some interpretations for this discrepancy between hot and cold-dust evolution may then be given: either the reduced star-formation at low redshifts is unable to heat enough quantities of dust for it to dominate at 3.3--4.2\,$\mu$m, thus becoming diluted in the much brighter stellar flux; or dust may be, with decreasing redshift, gradually farther from UV/optical sources (e.g., blown by strong radiation fields from an active nucleus or stellar winds), thus decreasing the relative contribution of the hottest spectral dust-components; or, in the case of the AGN population, different evolution of the heating mechanisms at play --- AGN activity dominating the hot-dust regime versus star-formation dominating the cold-dust regime --- in AGN hosts.

Table~\ref{tab:fracld} shows, for each redshift interval, the contributions of each of the galaxy populations to the overall dust luminosity density at rest-frames 3.3 and 4.2\,$\mu$m. These values refer to contributions down to the adopted luminosity completeness cut.

\begin{deluxetable}{ccrrrr}
\tabletypesize{\normalsize}
\tablecaption{The contribution of the different galaxy samples to the hot-dust (top rows) and PAH (bottom rows) luminosity densities.\label{tab:fracld}}
\tablewidth{0pt}
\tablehead{
\colhead{} & \colhead{$z_{\rm BIN}$} & \colhead{EARLY} & \colhead{LATE} & \colhead{STARB.} & \colhead{AGN}\\
\colhead{} & \colhead{} & \colhead{[\%]} & \colhead{[\%]} & \colhead{[\%]} & \colhead{[\%]}
}
\startdata
Dust & $0.05<z<0.19$ & 2 & 23 & 72 & 3 \\
& $0.21<z<0.52$ & 6 & 28 & 47 & 19 \\
& $0.52<z<0.94$ & 3 & 19 & 45 & 33 \\
\hline\\
PAH & $0.05<z<0.19$ & 6 & 18 & 76 & 0 \\
& $0.21<z<0.52$ & 12 & 35 & 53 & 0 \\
& $0.52<z<0.94$ & 15 & 28 & 56 & 1
\enddata
\tablecomments{The values are approximated to unit.\vspace{0.25cm}}
\end{deluxetable}

\section{Conclusions} \label{sec:conc}

In this work, the properties of hot dust ($\sim$690\,K) emission was explored using observations tracing rest-frame 4.2\,$\mu$m. This wavelength is also used to infer the dust contribution at rest-frame 3.3\,$\mu$m, and help assessing 3.3-PAH strength. Our approach considers stellar, dust, and PAH emissions separately, as well as the separation of the IR galaxy population into early, late, starburst, and IR-selected AGN. This allows the evaluation of the IR luminosity functions depending on galaxy-type and distance, as well as to estimate how much dust is contributing to the IR emission. We conclude:

\begin{itemize}

\item[-] Evolution with redshift of the hot-dust luminosity densities resembles that of cold-dust, but it drops more steeply (at least 0.5\,dex more) with decreasing redshift. The discrepancy is larger if one attempts to correct for sample incompleteness. The reason for the discrepancy is probably a combination of: with decreasing redshift, the star-formation becomes gradually unable to heat dust to such high temperatures for it to significantly contribute to the galaxy SED at such short wavelengths; dust is gradually located at increasingly larger distances from the heating sources (stars or active nuclei); and/or distinct evolution of different heating mechanisms at play in AGN hosts. This trend gives support to the scenario where galaxy obscuration increases with redshift, as already proposed in the literature (Section~\ref{sec:ldz}).

\item[-] While the average dust-to-total luminosity ratio is observed to increase with redshift, the average PAH-to-total luminosity ratio is consistent with being constant. The hot-dust-to-PAH luminosity ratio also depends on galaxy spectral type. A plausible explanation may be related to the findings of \citet{Mentuch09}, who observe increased extra emission at 2--5\,$\mu$m with increasing star-formation rates (Section~\ref{sec:dlfs}).

\item[-] Hot-dust and PAH LDFs of the non-AGN populations show that the hot-dust and PAH emission in the most luminous galaxies at 1.6\,$\mu$m (i.e., the most massive) decreases faster (with decreasing redshift) than in less luminous galaxies (Section~\ref{sec:dlfs}).

\item[-] Down to our luminosity limit, the AGN population comprises only $\lesssim3\%$ of the total galaxy population at $z<0.94$, but its contribution to the overall hot-dust luminosity density increases from 3\% at $0.05 < z < 0.19$ to 33\% at $0.52 < z < 0.94$, and it clearly dominates the bright-end of the total hot-dust LDFs at $0.52 < z < 0.94$ (Section~\ref{sec:dlfs}). The main driver for this is the high dust-to-total luminosity ratio at rest-frame 4.2\,$\mu$m found in IR-selected AGN: $\sim$0.8 (Section~\ref{sec:d2t}).

\item[-] At $\rm{M}_{1.6}>-25$, there is an increase of the dust-to-total and PAH-to-total luminosity ratios with decreasing luminosity, but deeper data may be required to confirm this result (Section~\ref{sec:d2t}).
\end{itemize}

\acknowledgments

The authors thank the COSMOS team for providing the photometric and redshift catalogs which make the base of this work. HM acknowledges the support from Funda\c{c}\~{a}o para a Ci\^{e}ncia e a Tecnologia through the scholarship SFRH/BD/31338/2006 and CONICyT-ALMA by the post-doc scholarship under the project 31100008. HM and JA acknowledge support from Funda\c{c}\~{a}o para a Ci\^{e}ncia e a Tecnologia through the projects PTDC/CTE-AST/105287/2008, PEst-OE/FIS/UI2751/2011 and PTDC/FIS-AST/2194/2012. HM acknowledges the support by UCR while visiting Dr. Bahram Mobasher as a visitor scholar. HM acknowledges the frequent use of C, Topcat, Supermongo, and Python.

\begin{appendix}

\section{A. Photometric redshifts for IR-selected AGN} \label{sec:zphagn}

The photometric redshift ($z_{\rm phot}$) analysis adopted in this work (and done in I09) is based on 0.15--8\,$\mu$m photometry. However, AGN activity may induce emission in this spectral regime, thus affecting the analysis \citep[][S11]{RowanRobinson08}. Here, we test the $z_{\rm phot}$ quality obtained for sources in our IR-selected AGN sample.

For the purpose of quality check we adopt the normalised median absolute deviation \citep[NMAD, by definition $\sigma_{\rm NMAD}=1.48\times \rm{median}(|z_{\rm spec}-z_{\rm phot}|/(1+z_{\rm spec}))$,][]{Hoaglin83}, which, for a Gaussian distribution, is directly comparable to the $\sigma_{\Delta z/(1+z_{\rm spec})}$ value quoted in other papers. Also, the fraction of outliers (sources with $|z_{\rm spec}-z_{\rm phot}|/(1+z_{\rm spec})>0.15$) is denoted by $\eta$. The discussion below only considers sources with $i^+<25\,$AB.

Among the IR AGN sample, there are 605 sources with good quality spectra. Comparing the estimated spectroscopic redshifts with the photometric redshifts estimated by I09 ($z_{\rm I09}$), the quality achieved is $\sigma_{\rm NMAD}=0.123$, with $\eta=39\%$. \citeauthor{Ilbert09} also computed in parallel the best-fit solution using AGN-like templates (those considered by S11). If one adopts these estimates instead ($z_{\rm AGN}$), the quality is improved to $\sigma_{\rm NMAD}=0.036$ and $\eta=23\%$. This is similar to what is achieved when applying the \citet{Salvato09} method to \emph{Chandra}-detected sources, even though plagued by a higher fraction of outliers. One can further improve these results by adopting $z_{\rm I09}$ or $z_{\rm AGN}$ solutions depending on which presents the lowest $\chi^2$ value. This procedure assigns the $z_{\rm AGN}$ solution to 66\% of the KI-selected sample with spectra, and reduces both the $\sigma_{\rm NMAD}$ and $\eta$ values to $\sigma_{\rm NMAD}=0.021$ and $\eta=18\%$.

\begin{figure}
\plottwo{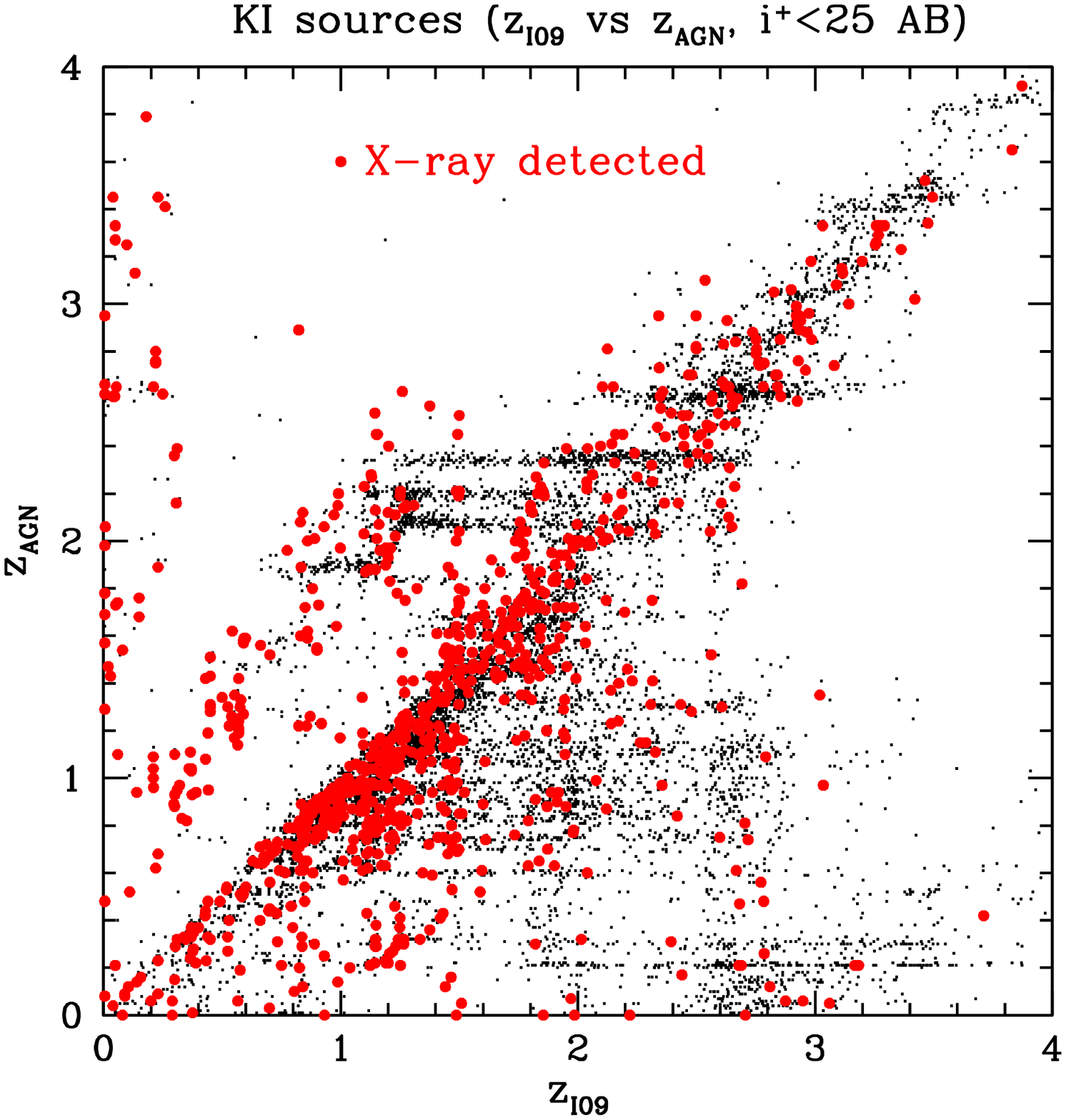}{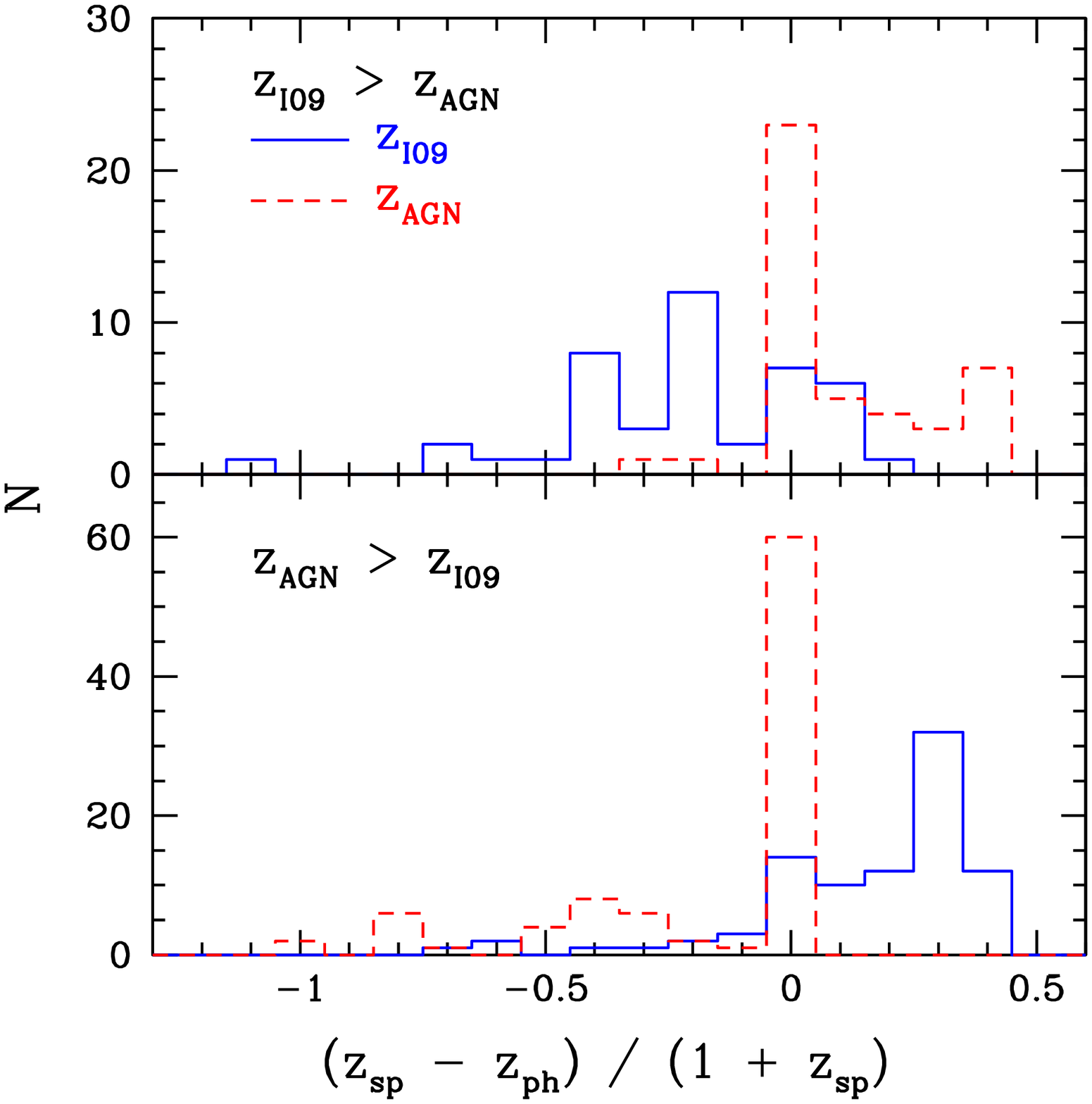}
\caption{Comparing photometric estimates obtained with non-AGN templates ($z_{\rm I09}$) and AGN-like templates ($z_{\rm AGN}$). The left hand-side panel compares the two estimates directly. Red dots show sources with an X-ray detection. The right hand-side panel compares the uncertainty of the two photometric estimates compared to spectroscopic measurement for two samples where $z_{\rm I09}$ and $z_{\rm AGN}$ strongly disagree (see text).\label{fig:zagni09}}
\end{figure}

Figure~\ref{fig:zagni09} compares directly $z_{\rm I09}$ and $z_{\rm AGN}$ solutions for IR-selected AGN sources. One can identify three regions in the plots where the two model sets provide quite distinct solutions: (A) sources with $z_{\rm I09}\gtrsim1$ and $z_{\rm AGN}\lesssim0.5$; (B) sources with $z_{\rm I09}\gtrsim1.7$ and $0.6\lesssim z_{\rm AGN}\lesssim1.5$; (C) sources with $0.8\lesssim z_{\rm I09}\lesssim1.8$ and $z_{\rm AGN}\gtrsim1.8$. In these regions, both AGN and I09 solutions present on average $\chi^2\sim1$ (meaning both provide a good fit to the observed data), hence the $\chi^2$ check referred above is irregular in these specific samples.

In region A there are 18 sources with spectroscopy, all but three fall at $z_{\rm spec}<1$ (only one at $z_{\rm spec}<0.5$). The X-ray detection rate (58 sources among 1724) is rather low for a $z<0.5$ AGN population, but acceptable for a higher redshift population including both AGN and non-AGN sources. Not only the XMM-\emph{Newton} and \emph{Chandra} coverages gradually miss more AGN sources with increasing redshift, but the reader should also be reminded that KI AGN-selection is quite contaminated by non-AGN sources at $z\gtrsim2.5$, where 63\% of the A sample likely falls (considering the photometric redshifts from I09). In fact, considering S11 redshift estimates ($z_{\rm S11}$) for the 58 X-ray detected sources, only two of them fall at $z<1$. These evidences support the scenario where sample A is mostly composed by $z>1$ sources, and we thus assign the photometric redshifts from I09 to sources in sample A. We note however, that 37 of these sources (64\%) do require AGN-like templates (S11). The difference between $z_{\rm S11}$ and $z_{\rm AGN}$ is likely not just due to variability (about 30\% of those sources are non-variable, i.e., variability below 0.25; S11), but rather a result from not considering luminosity priors (as used in S11) when computing $z_{\rm AGN}$.

In sample B (selected as sources with $z_{\rm AGN}>0.5$, $z_{\rm I09}>z_{\rm AGN}$, and $|z_{\rm I09}-z_{\rm AGN}|/(1+z_{\rm AGN})>0.2$) we do not find such a clear trend, with both spectroscopic measurements and $z_{\rm S11}$ estimates showing equal evidence for a sample composed by sources at $z<1.5$ and $z>1.5$. The upper histogram in Figure~\ref{fig:zagni09} shows the redshift uncertainty distribution by comparing $z_{\rm I09}$ and $z_{\rm AGN}$ to $z_{\rm sp}$ for sample B (where $z_{\rm I09}>z_{\rm AGN}$). The peak of the $z_{\rm AGN}$ distribution at zero is dominated by sources which already show $\chi^2_{\rm AGN}<\chi^2_{\rm I09}$. Hence, we do not modify the assigned photometric redshifts resulting from the comparison between $\chi^2$ values.

Sources found in region C (selected as $z_{\rm AGN}>0.5$, $z_{\rm AGN}>z_{\rm I09}$, and $|z_{\rm I09}-z_{\rm AGN}|/(1+z_{\rm AGN})>0.1$) mostly support the $z_{\rm AGN}$ solution. The peak in the uncertainty distribution in the lower histogram in Figure~\ref{fig:zagni09} is dominated by sources with $\chi^2_{\rm I09}/\chi^2_{\rm AGN}>0.7$, while sources with a $z_{\rm AGN}$ uncertainty of $\sim$-0.4 mostly have $\chi^2_{\rm I09}/\chi^2_{\rm AGN}<0.7$ and the $z_{\rm I09}$ uncertainty is around zero. The remainder sources with large $z_{\rm AGN}$ uncertainties have $z_{\rm S11}$ estimates and most show variability.

\begin{figure}
\epsscale{0.5}
\plotone{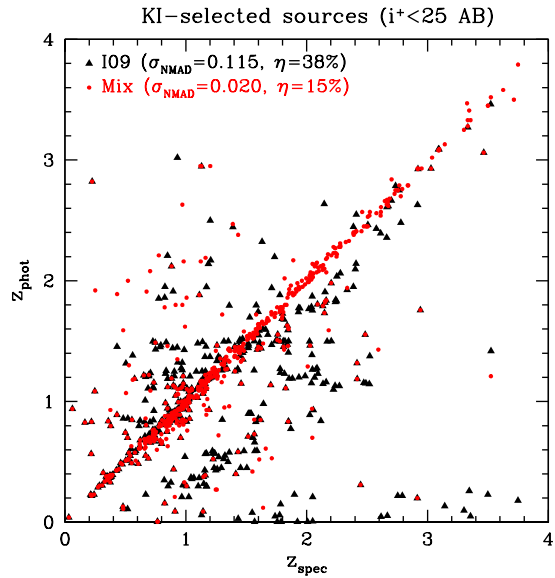}
\caption{Testing the match between spectroscopic (x-axis) and photometric (y-axis) redshift estimates. Black triangles consider only the $z_{\rm phot}$ estimates from I09, while red circles consider a mixture between $z_{\rm phot}$ estimates resulting either from AGN-model fit or non-AGN (I09) model fit as described in the text.\label{fig:zspph}}
\end{figure}

Hence, for the IR-selected AGN sample, we adopt the $z_{\rm AGN}$ solution for sources with $z_{\rm AGN}>0.5$ and: $z_{\rm AGN}>z_{\rm I09}$ and $\chi^2_{\rm I09}/\chi^2_{\rm AGN}>0.7$; or $z_{\rm AGN}<z_{\rm I09}$ and $\chi^2_{\rm I09}/\chi^2_{\rm AGN}>1$. The remainder of the population is assigned the $z_{\rm I09}$ solution. The photometric redshift quality achieved for the IR-selected AGN sample when compared to spectroscopic estimates is $\sigma_{\rm NMAD}=0.020$ and $\eta=15\%$ (Figure~\ref{fig:zspph}). The reader should recall that these $z_{\rm phot}$ estimates become obsolete when a higher priority redshift estimate is available (such as spectroscopic or S11 $z_{\rm phot}$ estimates, see Section~\ref{sec:reds}). For instance, by assuming the S11 $z_{\rm phot}$ estimates, which take into account luminosity-priors and the effect of variability, the photometric redshift quality is improved to $\sigma_{\rm NMAD}=0.015$ and $\eta=10\%$.\\

\end{appendix}

\twocolumngrid


\begin{thebibliography}{}
\bibitem[Aird et al.(2010)]{Aird10} Aird, J., et al.\ 2010, \mnras, 401, 2531 
\bibitem[Assef et al.(2011)]{Assef11} Assef, R.~J., et al.\ 2011, \apj, 728, 56 
\bibitem[Barlow et al.(2010)]{Barlow10} Barlow, M, et al. 2010, \aap, 518, 138
\bibitem[Blain et al.(1999)]{Blain99} Blain, A. W.; Smail, Ian; Ivison, R. J. \& Kneib, J.-P., \mnras, 302, 632
\bibitem[Bouwens et al. (2009)]{Bouwens09} Bouwens, R. J., et al. 2009, \apj, 705, 936
\bibitem[Brusa et al.(2010)]{Brusa10} Brusa, M., et al. 2010, \apj, 716, 348
\bibitem[Bruzual \& Charlot (2003)]{BruzualCharlot03} Bruzual, G. \& Charlot, S. 2003, \mnras, 344, 1000
\bibitem[Buat et al.(2005)]{Buat05} Buat, V., et al 2005, \apj, 619, 51
\bibitem[Calzetti et al.(2007)]{Calzetti07} Calzetti, D., Kennicutt, R.~C., Engelbracht, C.~W., et al.\ 2007, \apj, 666, 870 
\bibitem[Chary \& Pope(2010)]{CharyPope10} Chary, R.-R., \& Pope, A.\ 2010, arXiv:1003.1731 
\bibitem[Choi et al.(2006)]{Choi06} Choi, P.~I., Yan, L., Im, M., et al.\ 2006, \apj, 637, 227
\bibitem[Clements et al.(2010)]{Clements10} Clements, D.~L., Rigby, E., Maddox, S., et al.\ 2010, \aap, 518, L8
\bibitem[Croom et al.(2004)]{Croom04} Croom, S.~M., Smith, R.~J., Boyle, B.~J., Shanks, T., Miller, L., Outram, P.~J., \& Loaring, N.~S.\ 2004, \mnras, 349, 1397
\bibitem[da Cunha et al.(2008)]{daCunha08} da Cunha, E., Charlot, S., \& Elbaz, D.\ 2008, \mnras, 388, 1595 
\bibitem[Dasyra et al.(2009)]{Dasyra09} Dasyra, K.~M., Yan, L., Helou, G., et al.\ 2009, \apj, 701, 1123 
\bibitem[Donley et al.(2012)]{Donley12} Donley, J.~L., Koekemoer, A.~M., Brusa, M., et al.\ 2012, \apj, 748, 142
\bibitem[Drory et al.(2009)]{Drory09} Drory, N., et al. 2009, \apj, 707, 1595
\bibitem[Dunne et al.(2011)]{Dunne11} Dunne, L., Gomez, H.~L., da Cunha, E., et al.\ 2011, \mnras, 417, 1510 
\bibitem[Fazio et al.(2004)]{Fazio04} Fazio, G.~G., Hora, J.~L., Allen, L.~E., et al.\ 2004, \apjs, 154, 10 
\bibitem[Ferrarotti \& Gail (2006)]{FerrarottiGail06} Ferrarotti, A. S. \& Gail, H.-P., 2006, \aap, 447, 553
\bibitem[Fu et al.(2010)]{Fu10} Fu, H., et al. 2010, \apj, 722, 653
\bibitem[Garn et al.(2010)]{Garn10} Garn, T., Sobral, D., Best, P.~N., et al.\ 2010, \mnras, 402, 2017 
\bibitem[Gehrz et al.(1989)]{Gehrz89} Gehrz, R. 1989, IAUS, 135, 445
\bibitem[Greve et al.(2004)]{Greve04} Greve, T.~R., Ivison, R.~J., Bertoldi, F., et al.\ 2004, \mnras, 354, 779
\bibitem[Greve et al.(2008)]{Greve08} Greve, T.~R., Pope, A., Scott, D., et al.\ 2008, \mnras, 389, 1489 
\bibitem[Hainline et al.(2011)]{Hainline11} Hainline, L.~J., Blain, A.~W., Smail, I., et al.\ 2011, \apj, 740, 96 
\bibitem[Hoaglin et al.(1983)]{Hoaglin83} Hoaglin, D.~C., Mosteller, F., \& Tukey, J.~W.\ 1983, Wiley Series in Probability and Mathematical Statistics, New York: Wiley, 1983, edited by Hoaglin, David C.; Mosteller, Frederick; Tukey, John W.,  
\bibitem[H\"onig \& Kishimoto (2010)]{HonigKishimoto10} H\"onig, S. F. \& Kishimoto, M. 2010, \aap, 523, 27
\bibitem[Hunt (2010)]{Hunt10} Hunt, L. K. 2010, MSAIS, 14, 78
\bibitem[Ilbert et al.(2009)]{Ilbert09} Ilbert, O., et al. 2009, \apj, 690, 1236
\bibitem[Imanishi et al.(2010)]{Imanishi10} Imanishi, M., Nakagawa, T., Shirahata, M., Ohyama, Y., \& Onaka, T.\ 2010, \apj, 721, 1233 
\bibitem[Ivison et al.(2005)]{Ivison05} Ivison, R.~J., Smail, I., Dunlop, J.~S., et al.\ 2005, \mnras, 364, 1025
\bibitem[Jacobs et al.(2011)]{Jacobs11} Jacobs, B. A., et al. 2011, \aj, 141, 110
\bibitem[Kartaltepe et al.(2010)]{Kartaltepe10} Kartaltepe, J.~S., Sanders, D.~B., Le Floc'h, E., et al.\ 2010, \apj, 709, 572 
\bibitem[Knobel et al.(2012)]{Knobel12} Knobel, C., Lilly, S.~J., Iovino, A., et al.\ 2012, \apj, 753, 121 
\bibitem[Lilly et al.(2009)]{Lilly09} Lilly, S. J., et al. 2009, \apjs, 184, 218
\bibitem[Magnelli et al.(2008)]{Magnelli08} Magnelli, B., Chary, R.~R., Pope, A., et al.\ 2008, \apj, 681, 258
\bibitem[Magnelli et al.(2009)]{Magnelli09} Magnelli, B., Elbaz, D., Chary, R.~R., et al.\ 2009, \aap, 496, 57 
\bibitem[Mauduit et al.(2012)]{Mauduit12} Mauduit, J.-C., Lacy, M., Farrah, D., et al.\ 2012, \pasp, 124, 714
\bibitem[Mentuch et al.(2009)]{Mentuch09} Mentuch, E., Abraham, R.~G., Glazebrook, K., et al.\ 2009, \apj, 706, 1020
\bibitem[Mentuch et al.(2010)]{Mentuch10} Mentuch, E., Abraham, R.~G., \& Zibetti, S.\ 2010, \apj, 725, 1971
\bibitem[Messias et al.(2012)]{Messias12} Messias, H., Afonso, J., Salvato, M., Mobasher, B., \& Hopkins, A.~M.\ 2012, \apj, 754, 120
\bibitem[Mortier et al.(2005)]{Mortier05} Mortier, A.~M.~J., Serjeant, S., Dunlop, J.~S., et al.\ 2005, \mnras, 363, 563
\bibitem[Nenkova et al.(2008)]{Nenkova08} Nenkova, M., Sirocky, M. M., Ivezi\'c, \v{Z}. \& Elitzur, M. 2008, \apj, 685, 147
\bibitem[Polletta et al.(2007)]{Polletta07} Polletta, M., et al. 2007, \apj, 663, 81
\bibitem[Popescu et al.(2011)]{Popescu11} Popescu, C.~C., Tuffs, R.~J., Dopita, M.~A., Fischera, J., Kylafis, N.~D., \& Madore, B.~F.\ 2011, \aap, 527, A109 
\bibitem[Pozzetti et al.(2010)]{Pozzetti10} Pozzetti, L., et al. 2010, \aap, 523, 13
\bibitem[Rho et al.(2008)]{Rho08} Rho, J. et al. 2008, \apj, 673, 271
\bibitem[Richards et al.(2005)]{Richards05} Richards, G.~T., et al.\ 2005, \mnras, 360, 839
\bibitem[Roche et al.(2003)]{Roche03} Roche, N. D., Dunlop, J., \& Almaini, O. 2003, \mnras, 346, 803
\bibitem[Rowan-Robinson et al.(2008)]{RowanRobinson08} Rowan-Robinson, 
M., Babbedge, T., Oliver, S., et al.\ 2008, \mnras, 386, 697
\bibitem[Sajina et al.(2009)]{Sajina09} Sajina, A., Spoon, H., Yan, L., et al.\ 2009, \apj, 703, 270
\bibitem[Salvato et al.(2009)]{Salvato09} Salvato, M., et al. 2009, \apj, 690, 1250
\bibitem[Salvato et al.(2011)]{Salvato11} Salvato, M., Ilbert, O., Hasinger, G., et al.\ 2011, \apj, 742, 61
\bibitem[Sargent et al.(2010)]{Sargent10} Sargent, B. A., et al. 2010, \apj, 716, 878
\bibitem[Saunders et al.(1990a)]{Saunders90a} Saunders, W. 1990, PhDT, 353
\bibitem[Saunders et al.(1990b)]{Saunders90b} Saunders, W.; Rowan-Robinson, M.; Lawrence, A.; Efstathiou, G.; Kaiser, N.; Ellis, R. S. \& Frenk, C. S. 1990, \mnras, 242, 318 
\bibitem[Schmidt (1968)]{Schmidt68} Schmidt, M. 1968, \apj, 151, 393
\bibitem[Scoville et al.(2007)]{Scoville07} Scoville, N., et al. 2007, \apjs, 172, 1
\bibitem[Scott et al.(2002)]{Scott02} Scott, S.~E., Fox, M.~J., Dunlop, J.~S., et al.\ 2002, \mnras, 331, 817
\bibitem[Siana et al.(2009)]{Siana09} Siana, B., Smail, I., Swinbank, A.~M., et al.\ 2009, \apj, 698, 1273 
\bibitem[Silva et al.(1998)]{Silva98} Silva, L., Granato, G.~L., Bressan, A., \& Danese, L.\ 1998, \apj, 509, 103 
\bibitem[Tielens (2011)]{Tielens11} Tielens, A. G. 2011, EAS, 46, 3
\bibitem[Trump et al.(2009)]{Trump09} Trump, J.~R., Impey, C.~D., Elvis, M., et al.\ 2009, \apj, 696, 1195 
\bibitem[Villar et al.(2008)]{Villar08} Villar, V., Gallego, J., P{\'e}rez-Gonz{\'a}lez, P.~G., et al.\ 2008, \apj, 677, 169
\bibitem[Werner et al.(2004)]{Werner04} Werner, M.~W., Roellig, T.~L., Low, F.~J., et al.\ 2004, \apjs, 154, 1 
\bibitem[Yamada et al.(2013)]{Yamada13} Yamada, R., Oyabu, S., Kaneda, H., et al.\ 2013, arXiv:1307.6356
\end{thebibliography}
\end{document}